\documentclass[12pt]{article}
\usepackage{epsfig, epsf, graphicx, subfigure}
\usepackage{pstricks, pst-node, psfrag}
\usepackage{amssymb,amsmath,bm}
\usepackage{verbatim,enumerate}
\usepackage{rotating, lscape}
\usepackage{setspace}
\usepackage[english]{babel}
\usepackage[sort&compress]{natbib}
\usepackage{multirow}
\usepackage{theorem}

\usepackage{appendix,hyperref}

\def\E{\mathrm{E}}
\def\d{\mathrm{d}}

\def\00{\mathrm{0}}
\def\UU{\mathbf{U}}

\def\vv{\mathcal{V}}
\def\ZZ{\mathbf{Z}}
\def\ZM{\mathcal{Z}}

\def\WW{\boldsymbol{W}}

\def\UU{\boldsymbol{U}}

\def\ww{\boldsymbol{w}}

\def\zz{\boldsymbol{z}}
\def\tht{\boldsymbol{\theta}}

\def\EE{\mathcal{E}}
\def\aa{\mathbf{a}}

\def\aa{\boldsymbol{\alpha}}
\def\bb{\boldsymbol{\beta}}
\def\ii{\boldsymbol{1}}

\def\OO{\boldsymbol{\Omega}}

\def\ss{\boldsymbol{s}}

\def\MM{\mathbf{M}}
\def\XX{\mathbf{X}}
\def\xx{\mathbf{x}}

\def\WW{\mathbf{W}}

\newtheorem{ass}{Assumption}
\newtheorem{prop}{Proposition}
\newtheorem{corol}{Corollary}

\newtheorem{lemma}{Lemma}

\begin{document}	
\thispagestyle{empty} \baselineskip=28pt \vskip 5mm
\begin{center} 
%{\Huge{\bf Cauchy convolution processes for the modeling of spatial extremes with local tail dependence}}
%\RH{\Huge{\bf Cauchy convolution processes for modeling spatial data with local tail dependence}}
\Huge{\bf A class of skew-multivariate distributions for spatial data}
\end{center}
	
\baselineskip=20pt \vskip 10mm

\begin{center}\large
Pavel Krupskii\footnote[1]{\baselineskip=10pt University of Melbourne, Parkville, Victoria, 3010, Australia. E-mail: pavel.krupskiy@unimelb.edu.au.} 
\end{center}

%\begin{document}

\begin{center}
	{\large{\bf Abstract}}
\end{center}

This paper introduces a class of copula models for spatial data, based on multivariate Pareto-mixture distributions. We explore the tail properties of these models, demonstrating their ability to capture both tail dependence and asymptotic independence, as well as the tail asymmetry frequently observed in real-world data. The proposed models also offer flexibility in accounting for permutation asymmetry and can effectively represent both the bulk and extreme tails of the distribution. We consider special cases of these models with computationally tractable likelihoods and present an extensive simulation study to assess the finite-sample performance of the maximum likelihood estimators. Finally, we apply our models to analyze a temperature dataset, showcasing their practical utility.

\section{Introduction}

Modeling spatial data is a challenging task and it requires flexible models that can capture complex dependence structures often observed in the real-world datasets. Traditional approaches in geostatistics typically rely on the assumption of multivariate normality, as Gaussian random fields are computationally tractable and can be conveniently parameterized using a covariance function to represent spatial dependence \citep{Gneiting2002, Gneiting.Genton.ea2007}. However, this assumption often fails to hold in many practical scenarios. In particular, the Gaussian models are not suitable for data with skewed or heavy-tailed marginals and they cannot adequately capture tail dependence or asymmetric dependence. As a result, more flexible modeling approaches are needed to address these limitations and better reflect the underlying data characteristics.

To develop models with flexible marginals, trans-Gaussian random fields can be employed \citep{Xu.Genton2017}. For constructing non-Gaussian random fields, scale mixture models \citep{Ma2009a}, log-skew elliptical distributions \citep{Marchenko.Genton2010}, skew-normal distributions \citep{Genton.Zhang2012, Rimstad.Omre2014}, as well as $t$-distributions and skew-$t$ distributions \citep{Roislien.Omre2006, Bevilacqua.ea2021} can be utilized. These approaches offer greater flexibility in capturing the complex features of real-world data, such as skewness and heavy tails.

To combine flexibility in both the marginals and the joint distribution, copula models have gained popularity in spatial data modeling. A copula is a function that links univariate marginal distributions to form a multivariate cumulative distribution function (CDF). \cite{Sklar1959} demonstrated that for any multivariate CDF $F_{1:d}$ with marginals $F_1,\ldots,F_d$ there exists a copula $C_{1:d}: [0,1]^d \to [0,1]$ such that
$$
F_{1:d}(z_1, \ldots, z_d) = C_{1:d}\{F_1(z_1), \ldots, F_d(z_d)\},
$$
and this copula is unique if the marginals are continuous. 

For a pair of variables $(Z_1, Z_2)$ with continuous marginal distributions $F_1, F_2$ and copula $C$, the lower and upper tail dependence coefficients, $\lambda_L$ and $\lambda_U$ are defined as
\begin{eqnarray*}
\lambda_L :=& \lim_{q \to 0} \Pr\{Z_1 < F_1^{-1}(q) | Z_2 < F_2^{-1}(q)\} =& \lim_{q \to 0} \frac{C(q,q)}{q}\,,\\
\lambda_U :=& \lim_{q \to 0} \Pr\{Z_1 > F_1^{-1}(q) | Z_2 > F_2^{-1}(q)\} =& \lim_{q \to 0} \frac{\bar C(q,q)}{q}\,,
\end{eqnarray*}
where $\bar C(u_1,u_2) = -1 + u_1 + u_2 + C(1-u_1, 1-u_2)$ is the survival copula. If $\lambda_L > 0$ or $\lambda_U > 0$, the pair $(Z_1, Z_2)$ is said to exhibit the lower (upper) tail dependence which indicating strong dependence in the joint lower or upper tail. Moreover, if $C(u_1, u_2) = \bar C(u_1, u_2)$ for all $(u_1, u_2) \in (0,1)^2$, the copula $C$ is called radially symmetric,  implying a symmetric tail structure.

While models based on the multivariate normality assumption result in tail independence and radial symmetry, copulas provide a powerful framework for constructing flexible distributions with various tail dependence structures.  Examples include v-transformed copulas \citep{Bardossy.Li2008, Bardossy2011} and vine copulas \citep{Graler.Pebesma2011, Graler2014}. However, these models often lack interpretability. Copulas based on non-monotone transformations of normal vectors \citep{Quessy.ea2015, Quessy.ea2016} can capture radial asymmetry, but they imply tail independence and become computationally intractable in very high dimensions unless all non-centrality parameters are assumed equal, which in turn reduces model flexibility. In contrast, factor copula models are more parsimonious and easier to interpret. Factor copula models have been successfully adapted for spatial data modeling \citep{Krupskii.Genton2017, Krupskii.Huser.ea2018}, providing a balance between flexibility and interpretability.

While factor copula models are interpretable, they assume the existence of one or more factors that influence all spatial locations simultaneously. As a result, these models cannot fully capture the independence typically observed at very large distances. In fact, most factor copula models for spatial data proposed in the literature imply tail dependence—specifically, $\lambda_L > 0$ or $\lambda_U > 0$ for any pair of variables—irrespective of the distance between the corresponding locations. This assumption is often unrealistic in many applications. To address this, \cite{Wadsworth.Tawn2012, Huser.Opitz.ea2017,Huser.Wadsworth2019} introduced models for spatial data that allow for a transition between tail dependence and tail independence, depending on the chosen model parameters. Further advancements by \cite{Krupskii.Huser2022, Krupskii.Huser2024} have led to models that permit tail dependence at small distances and tail independence (or even full independence) at larger distances. However, these models become computationally intractable in very high dimensions.

Furthermore, many copula models for spatial data presented in the literature assume permutation symmetry, which can be a restrictive assumption in certain applications. A copula $C$ that links a pair of variables $(Z_1, Z_2)$ is said to be permutation symmetric if $C(u_1, u_2) = C(u_2, u_1)$ for all $(u_1, u_2) \in (0,1)^2$. However, permutation asymmetric models are often better equipped to handle data with complex extremal dependence \citep{Beranger.Padoan.ea2017}. Examples of permutation asymmetric models for multivariate extremes include extremal skew normal and extremal skew-$t$ models \citep{Padoan2011,Beranger.Padoan.ea2019,Beranger.Stephenson.ea2021}. While these models are suitable for capturing the limiting behavior of extreme events, they may not provide accurate fits for other types of data.

Classical skew-multivariate models include the skew-normal and skew-$t$ distributions \citep{Azzalini.DallaValle1996, Azzalini.Capitanio2003}, as well as several extended families \citep{Arellano.Genton2010}. However, these models can be computationally demanding due to the complex geometry of their likelihood functions, although more stable algorithms have been developed for skew-$t$ copula models \citep{Yoshiba2018, Deng.Smith.ea2024}. While these models can provide a good fit in the bulk of the distribution, they tend to lack flexibility in the tails. Specifically, the skew-normal copula cannot capture tail dependence, and the skew-$t$ copula assumes both lower and upper tail dependence, making it unsuitable for modeling data that exhibit tail dependence in only one of the tails.

In this paper, we study a class of copula models for spatial data based on skew-multivariate distributions that provide flexibility both in the bulk and in the tails of the distribution. Specifically, we focus on a Pareto-mixture random process of the form
\begin{equation}
    \label{main-eq0}
    \XX(\ss) = P^{1/\nu(\ss)} \ZZ(\ss),
\end{equation}
where $\ZZ(\ss)$ is a spatial process (e.g., a Gaussian process), $P > 0$ is a continuous random variable independent of both $\ZZ(\ss)$ and the spatial location $\ss$, and $\nu(\ss)$ is a positive, spatially varying parameter. We assume that $P$ follows a distribution with a Pareto-type tail, with the probability density function (PDF) $$f_P(z) = \xi_Pz^{-2} + o\left(z^{-2}\right) \quad \text{as} \quad z \to \infty,$$ for some constant $\xi_P > 0$. 

To construct the corresponding copula, we restrict model \eqref{main-eq0} to a finite set of spatial locations $\ss_1, \ldots, \ss_d \in \mathbb{R}^p$. Let $Z_i = \ZZ(\ss_i)$ and $\nu_i = \nu(\ss_i)$, and define the vector 
\begin{equation}
    \label{main-eq}
    \XX = (X_1, \ldots, X_d)^{\top}, \quad X_i = \XX(\ss_i) = Z_i P^{1/\nu_i}, \quad i \in \{1, \ldots, d\}.
\end{equation}

Our goal is to characterize the copula $C_{\XX}$ of $\XX$. In this framework, the distribution of $\ZZ = (Z_1, \ldots, Z_d)^{\top}$ --- such as a multivariate normal or skew-normal distribution --- governs the bulk behavior of the joint distribution of $\XX$ and can be parameterized using a covariance function to capture spatial dependence, while the Pareto factor $P$ controls the tail behavior. In addition, permutation asymmetry is governed by the positive parameters $\nu_1, \ldots, \nu_d$.

We examine several notable special cases in which the full likelihood has a simple closed-form expression, allowing for straightforward inference. We also illustrate that these models can accommodate spatial data exhibiting highly complex dependence structures.

%the random vector $\ZZ = (Z_1,\ldots,Z_d)^{\top}$ follows a multivariate distribution  with CDF $F_{\ZZ}$. In this framework, the distribution of $\ZZ$ --- such as multivariate normal or skew-normal distribution --- controls the bulk of the joint distribution of $\XX$, while the Pareto factor $P$ controls its tails. Furthermore, permutation asymmetry is controlled by the positive parameters $\nu_1, \ldots, \nu_d$.

%Specifically, we focus on Pareto-mixtures of the form:
%\begin{equation}
%    \label{main-eq}
%    \XX = (P^{1/\nu_1} Z_1, \ldots, P^{1/\nu_d} Z_d)^{\top},
%\end{equation}
%where $P > 0$ is a continuous random variable following a distribution with a Pareto tail, characterized by the probability density function (PDF) $f_P(z) = \xi_Pz^{-2} + o\left(z^{-2}\right)$ as $z \to \infty$, where $\xi_P > 0$ is some constant and the random vector $\ZZ = (Z_1,\ldots,Z_d)^{\top}$ follows a multivariate distribution  with CDF $F_{\ZZ}$. In this framework, the distribution of $\ZZ$ --- such as multivariate normal or skew-normal distribution --- controls the bulk of the joint distribution of $\XX$, while the Pareto factor $P$ controls its tails. Furthermore, permutation asymmetry is controlled by the positive parameters $\nu_1, \ldots, \nu_d$.

The remainder of this paper is organized as follows. Section \ref{sec-model} introduces the modeling framework and examines the dependence properties of the proposed class of models, while Section \ref{sec-spec-case} presents several noteworthy special cases. Section \ref{sec-infer} covers maximum likelihood inference for these models, including a discussion of special cases with tractable likelihoods. Section \ref{sec-sim} evaluates the finite sample performance of the maximum likelihood estimators. In Section \ref{sec-empstudy}, we apply the proposed models to analyze a temperature dataset, and Section \ref{sec-conc} concludes with a discussion.

\section{Pareto-mixture distributions and their tail properties}
\label{sec-model}

\subsection{The model and its copula}
\label{subsec-model}

We consider the copula $C_{\XX}$ and its density $c_{\XX}$ of the vector $\XX$ as defined in \eqref{main-eq}. Throughout the paper, we assume that $Z_1, \ldots, Z_d$ are continuous random variables, so that $C_{\XX}$ is uniquely defined:
\begin{equation}
    \label{eq-copmodel}
    \begin{aligned}
    C_{\XX}(u_1, \ldots, u_d) &=& F_{\XX}\{F_{X_1}^{-1}(u_1),\ldots, F_{X_d}^{-1}(u_d)\},\\
    c_{\XX}(u_1, \ldots, u_d) &=& \frac{f_{\XX}\{F_{X_1}^{-1}(u_1),\ldots, F_{X_d}^{-1}(u_d)\}}{\prod_{i=1}^d f_{X_i}\{F_{X_i}^{-1}(u_i)\}},
    \end{aligned}
\end{equation}
where $F_{\XX}$ and $f_{\XX}$ are the joint CDF and PDF of $\XX$, and $F_{X_i}$ and $f_{X_i}$ are the marginal CDF and PDF of $X_i$, respectively. From \eqref{main-eq}, we find:
\begin{eqnarray*}
     F_{\XX}(x_1, \ldots, x_d) & = &\int_0^{\infty} F_{\ZZ}\left(x_1 y^{-1/\nu_1}, \ldots, x_d y^{-1/\nu_d}\right) f_P(y) \d y,\\
     f_{\XX}(x_1, \ldots, x_d) & = &\int_0^{\infty} f_{\ZZ}\left(x_1 y^{-1/\nu_1}, \ldots, x_d y^{-1/\nu_d}\right) y^{-\sum_{i=1}^d 1/\nu_i}f_P(y) \d y,
\end{eqnarray*}
and
\begin{eqnarray*}
    F_{X_i}(x_i) & = &  \int_0^{\infty} F_{Z_i}\left(x_i y^{-1/\nu_i}\right) f_P(y) \d y,\\
    f_{X_i}(x_i) & = &  \int_0^{\infty} f_{Z_i}\left(x_i y^{-1/\nu_i}\right) y^{-1/\nu_i}f_P(y) \d y.
\end{eqnarray*}
Here, $F_{\ZZ}$ and $F_{Z_i}$ are the joint CDF of $\ZZ$ and marginal CDF of $Z_i$, respectively. The distribution of $\ZZ$ can be chosen to capture spatial dependence, and many models commonly used in spatial statistics --- such as the multivariate normal and skew-normal distributions --- have closed-form expressions for their PDFs. Consequently, the density $f_{\XX}$, along with the marginal CDFs, can be expressed as a one-dimensional integral that can be evaluated numerically in the general case. This enables the use of the maximum likelihood approach to estimate the model parameters. In Section 3, we explore special cases of the model in \eqref{main-eq} where $f_{\XX}$ and the marginal CDFs are available in closed form, eliminating the need for numerical integration. This makes the corresponding models computationally feasible in high dimensions.

\subsection{Tail properties}
\label{subsec-tailprop}

While $F_{\ZZ}$ controls the bulk of the distribution of the vector $\XX$ as defined in \eqref{main-eq}, the Pareto factor $P$ controls its tail behavior. We now study the tail properties of this vector. First, we define the limiting extreme-value (EV) copula of $\XX$.  Let $\{(X_{j1}, \ldots, X_{jd})^{\top}\}_{j=1}^n$ be $n$ i.i.d. copies of the vector $\XX$, and define the vector of component-wise maxima as $\MM_n = (M_{n1}, \ldots, M_{nd})^{\top}$ where $M_{ni} = \max_{j=1,\ldots,n} X_{ij}$, $i \in \{ 1,\ldots, d\}$. Assume there exist sequences of normalizing constants $\mathbf{a_n} = (a_{n1},\ldots,a_{nd})^{\top} \in \mathbb{R}^d_+$ and $\mathbf{b_n} = (b_{n1},\ldots, b_{nd})^{\top} \in \mathbb{R}^d$ such that $\mathbf{a_n^{-1}}(\MM_n - \mathbf{b_n}) \to_d \XX^*$ where the vector $\XX^*$ has the CDF $F_{\XX^*}$ with nondegenerate marginals. The EV copula $C_{\XX}^{\mathrm{EV}}$ is then defined as the copula of the limiting distribution $F_{\XX^*}$, and can be expressed as:
$$
C_{\XX}^{\mathrm{EV}}(u_1,\ldots,u_d) = \lim_{n \to \infty} C_{\XX}^n(u_1^{1/n},\ldots,u_d^{1/n}), \quad (u_1, \ldots, u_d)^{\top} \in [0,1]^d. 
$$
The EV copula $C_{\XX}^{\mathrm{EV}}$ can be also expressed in terms of the stable (upper) tail dependence function $\ell_{\XX}: \mathbb{R}^d_+ \to \mathbb{R}_+$ as:
$$
C_{\XX}^{\mathrm{EV}}(u_1, \ldots, u_d) = \exp\{-\ell_{\XX}(-\ln u_1, \ldots, -\ln u_d)\},
$$
where the stable tail dependence function is given by the following limit:
$$
\ell_{\XX}(w_1,\ldots,w_d) = \lim_{n\to\infty} n\left\{1-C_{\XX}\left(1-\frac{w_1}{n}, \ldots, 1 - \frac{w_d}{n}\right)\right\}.
$$
This function characterizes the limiting extremal behavior of the vector $\XX$ in the upper tail \citep{Segers2012, Ressel2013}. In particular, the quantity $\vartheta_{\XX} = \ell_{\XX}(1,\ldots,1)$, known as the extremal coefficient, ranges from 1 (complete or comonotonic dependence) to $d$ (asymptotic independence). To study the tail behavior of $\XX$ we impose the following assumption on the Pareto factor $P$.

\begin{ass}
\label{ass1}
Assume that $P > 0$ is a random variable with the PDF $f_P$ such that $\lim_{z \to \infty} z^2f_P(z) = \xi_P$ and $z^2 f(z) < C_P < \infty$ for some positive constants $\xi_P, C_P$ and for any $z > 0$. Further, assume that $F_{Z_i}(0) < 1$, $i \in \{1,\ldots,d\}$.
\end{ass}
    
The next proposition gives the stable tail dependence function of $\XX$ as defined in \eqref{main-eq}. 

\begin{prop}
    \label{prop1} Consider the random vector $\XX$ as given in \eqref{main-eq}. Let $F_{\ZZ}$ be the joint CDF of $\ZZ$ and assume there exists $\epsilon > 0$ such that the marginal survival function $\bar F_{Z_i}(z) = o(z^{-(1+\epsilon)\nu_i})$ for $i \in \{1,\ldots, d\}$ as $z \to \infty$. %Also assume that $P > 0$ is a continuous random variable with the PDF $f_P$ such that $\lim_{z \to \infty} z^2f_P(z) = 1$ and $z^2 f(z) < C_P < \infty$ for some constant $C_P$ and for any $z > 0$.
    Under these conditions and Assumption \ref{ass1}, the stable tail dependence function $\ell_{\XX}$ is given by
\begin{equation}
\label{eq-stdf}
\ell_{\XX}(w_1, \ldots, w_d) = \displaystyle\int_0^{\infty} \left[1 -  F_{\ZZ}\left\{\left(\frac{y}{w_1}\zeta_1\right)^{1/\nu_1},\ldots,\left(\frac{y}{w_d}\zeta_d\right)^{1/\nu_d}\right\}\right] \d y, 
\end{equation}
where $\zeta_i = \int_0^{\infty}\bar F_{Z_i}(y^{1/\nu_i}) \d y.$
\end{prop}

\textbf{Proof of Proposition \ref{prop1}:} Without loss of generality, we can assume $\xi_P = 1$. Let $\epsilon^* = \epsilon/(1+\epsilon)$. Define $$G_i(y,z) = \left\{1 - F_{Z_i}\left(\frac{z}{y^{1/\nu_i}}\right)\right\}f_P(y)\,.$$It follows that $\bar F_{X_i}(z) = \int_0^{\infty} G_i(y,z)\d y$  
and
\begin{align*}
\int_0^{z^{\nu_i\epsilon^*}}G_i(y, z) \d y &\leq \bar F_{Z_i}(z^{1-\epsilon^*}) = o\left\{(z^{-(1-\epsilon^*)(1+\epsilon)\nu_i}\right\} = o(z^{-\nu_i}),\\
\int_{z^{\nu_i\epsilon^*}}^{\infty}G_i(y, z) \d y &= \nu_iz^{\nu_i}\int_0^{z^{1-\epsilon^*}} \bar F_{Z_i}(y)y^{-\nu_i-1}f_P\left\{\left(\frac{z}{y}\right)^{\nu_i}\right\} \d y\\
& = z^{-\nu_i}\int_0^{z^{1-\epsilon^*}} \bar F_{Z_i}(y) \nu_i y^{\nu_i-1} \left(\frac{z}{y}\right)^{2\nu_i}f_P\left\{\left(\frac{z}{y}\right)^{\nu_i}\right\}\d y.
%& = z^{-\nu_i}\int_0^{z^{1-\epsilon^*}}  \bar F_{Z_i}(y) \nu_i y^{\nu_i-1} \d y +  o(z^{-\nu_i})\\
%&=\zeta_i z^{-\nu_i} + o(z^{-\nu_i}).
\end{align*}

For any $\varepsilon > 0$ we can find $M > 0$ such that
$$
\int_{M}^{\infty}\bar  F_{Z_i}(y) \nu_i y^{\nu_i-1}\d y < \varepsilon, 
$$
and then find $z(\varepsilon)$ such that
$$
\Delta_M = \Big|\int_0^M \bar F_{Z_i}(y) \nu_i y^{\nu_i-1} \left(\frac{z}{y}\right)^{2\nu_i}f_P\left\{\left(\frac{z}{y}\right)^{\nu_i}\right\}\d y - \int_0^M \bar F_{Z_i}(y) \nu_i y^{\nu_i-1} \d y\Big| < \varepsilon $$
for any $z > z(\varepsilon)$ by the dominated convergence theorem. It implies that
\begin{align*}    \Big|\int_0^{z^{1-\epsilon^*}} \bar F_{Z_i}(y) \nu_i y^{\nu_i-1} \left(\frac{z}{y}\right)^{2\nu_i}f_P\left\{\left(\frac{z}{y}\right)^{\nu_i}\right\}\d y  - \int_0^{\infty} \bar F_{Z_i}(y) \nu_i y^{\nu_i-1}\d y \Big| & \leq \Delta_M + \Delta_M^*, 
\end{align*}
where
\begin{align*}
    \Delta_M^* & = \Big|\int_M^{z^{1-\epsilon^*}} \bar F_{Z_i}(y) \nu_i y^{\nu_i-1} \left(\frac{z}{y}\right)^{2\nu_i}f_P\left\{\left(\frac{z}{y}\right)^{\nu_i}\right\}\d y  - \int_M^{\infty} \bar F_{Z_i}(y) \nu_i y^{\nu_i-1}\d y \Big|\\
    & \leq \Big|C_P\int_M^{\infty}\bar F_{Z_i}(y)\nu_iy^{\nu_i-1} \d y\Big| + \Big|\int_M^{\infty}\bar F_{Z_i}(y)\nu_iy^{\nu_i-1} \d y\Big| < (C_P+1)\varepsilon,
\end{align*}
and therefore 
$$
\int_0^{z^{1-\epsilon^*}} \bar F_{Z_i}(y) \nu_i y^{\nu_i-1} \left(\frac{z}{y}\right)^{2\nu_i}f_P\left\{\left(\frac{z}{y}\right)^{\nu_i}\right\}\d y \to \int_0^{\infty} \bar F_{Z_i}(y) \nu_i y^{\nu_i-1}\d y = \zeta_i \quad \text{if} \quad z \to \infty.
$$

It implies that $\bar F_{X_i}(z) = \zeta_iz^{-\nu_i} + o(z^{-\nu_i})$. Let $q_i = q_i(n) = (\zeta_i n/w_i)^{1/\nu_i}$, then $\bar F_{X_i}(q_i) = w_i/n + o(1/n)\,.$ Define $q_i^* = q_i^*(n) = F_{X_i}^{-1}\left(1-w_i/n\right)$. Since $\bar F_{X_i}(q_i) - \bar F_{X_i}(q_i^*) = o(1/n)$,
we have
$$
|F_{\XX}(q_1,\ldots,q_d) - F_{\XX}(q_1^*,\ldots,q_d^*)| \leq \sum_{i=1}^d |F_{X_i}(q_i) - F_{X_i}(q_i^*)| = o(1/n).
$$
It follows that 
\begin{align*}
\ell_{\XX}(w_1,\ldots, w_d) &= \lim_{n \to \infty} n\left\{1 - F_{\XX}(q_1^*,\ldots,q_d^*)\right\} 
%& = \lim_{n \to \infty} n\left\{1 - F_{\XX}(q_1,\ldots,q_d)\right\} + \lim_{n\to\infty} n\{F_{\XX}(q_1^*,\ldots,q_d^*) - F_{\XX}(q_1,\ldots,q_d)\}\\
 = \lim_{n \to \infty} n\left\{1 - F_{\XX}(q_1,\ldots,q_d)\right\} \\
&= \lim_{n \to \infty} n \int_0^{\infty}\left\{1 - F_{\ZZ}\left(\frac{q_1}{y^{1/\nu_1}},\ldots,\frac{q_d}{y^{1/\nu_d}}\right)\right\}f_P(y)\d y \\
& = \lim_{n \to \infty} \int_0^{\infty}\left[1 - F_{\ZZ}\left\{\left(\frac{y}{w_1}\zeta_1\right)^{1/\nu_1},\ldots,\left(\frac{y}{w_d}\zeta_d\right)^{1/\nu_d}\right\}\right] \left(\frac{n}{y}\right)^2f_P\left(\frac{n}{y}\right) \d y \\
& =  \int_0^{\infty}\left[1 - F_{\ZZ}\left\{\left(\frac{y}{w_1}\zeta_1\right)^{1/\nu_1},\ldots,\left(\frac{y}{w_d}\zeta_d\right)^{1/\nu_d}\right\}\right] \d y,
\end{align*}
where the last equality follows from the dominated convergence theorem. \hfill $\Box$

Note that the upper tail dependence coefficient for the pair $(X_i,X_k)^{\top}$, denoted $\lambda_{U}^{X_i,X_k}$, can be obtained from the corresponding stable tail dependence function  $\ell_{X_i,X_k}$ via
$\lambda_{U}^{X_i,X_k} = 2 - \ell_{X_i,X_k}(1,1)$. Proposition \ref{prop1} can therefore be used to compute the upper tail dependence coefficients for any pair of variables in model \eqref{main-eq}.

\begin{corol}
    \label{corol1}
    Consider the pair $(X_i, X_k)^{\top}$ for $i \neq k$, $i, k \in \{1,\ldots, d\}$. Under assumptions of Proposition \ref{prop1}, if $\Pr(Z_i > 0, Z_k > 0) > 0$, then the pair $(X_i, X_k)^{\top}$ exhibits upper tail dependence.
\end{corol}

\textbf{Proof of Corollary \ref{corol1}:} %Let $\ell_{X_i,X_k}$ denote the stable tail dependence of the pair $(X_i, X_k)^{\top}$. 
Recall that for each $l \in \{i,k\}$
$$
w_l  = \int_0^{\infty}\left[1 - F_{Z_l}\left\{\left(\frac{y}{w_l}\zeta_l\right)^{1/\nu_l}\right\}\right] \d y, \quad l \in \{i,k\},
$$
which follows from the definition of $\zeta_l$. 
We find for $w_i, w_k > 0$: 
$$
w_i + w_k - \ell_{X_i, X_k}(w_i,w_k) = \int_0^{\infty} \Pr\left\{Z_i > \left(\frac{y}{w_i}\zeta_i\right)^{1/\nu_i}, Z_k > \left(\frac{y}{w_k}\zeta_k\right)^{1/\nu_k}\right\} \d y > 0,
$$
so that $\ell_{X_i,X_k}(w_i,w_k) < w_i + w_k$ and $\lambda_U^{X_i,X_k} > 0$.  Hence, the pair $(X_1, X_k)^{\top}$ exhibits upper tail dependence. \hfill $\Box$

%As it follows from Proposition \ref{prop1}, if the Pareto factor $P^{1/\nu_i}$ dominates the corresponding component $Z_i$ in the upper tail, then the vector $\XX$ exhibits upper tail dependence. 
Next, we examine scenarios where  $\XX$ is asymptotically independent in the upper tail. 

\begin{prop}
    \label{prop2}
    Consider the $(i,k)$ margin of the random vector $\XX$ as defined in \eqref{main-eq}, and assume the conditions in Assumption \ref{ass1} are satisfied. 
    
    \noindent (a) Assume that there exists $\epsilon > 0$ such that the marginal survival functions satisfy $\bar F_{Z_i}(z) = o\left(z^{-(1+\epsilon)\nu_i}\right)$ and $\bar F_{Z_k}(z) = C_k z^{-\xi_k} + o\left(z^{-\xi_k}\right)$ as $z \to \infty$, where $\xi_k < \nu_k$ and $C_k > 0$ is a constant. Then, $(X_i, X_k)^{\top}$ exhibits no upper tail dependence. 
    
    \noindent (b) Assume that $\bar F_{Z_i}(z) = C_i z^{-\xi_i} + o\left(z^{-\xi_i}\right)$ and $\bar F_{Z_k}(z) = C_k z^{-\xi_k} + o\left(z^{-\xi_k}\right)$ as $z \to \infty$, where $\xi_i < \nu_i$, $\xi_k < \nu_k$ and $C_i, C_k > 0$ are constants. Additionally, assume that $(Z_i, Z_k)^{\top}$ exhibits no upper tail dependence. Then $(X_i, X_k)^{\top}$ also exhibits no upper tail dependence.
\end{prop}

\textbf{Proof of Proposition \ref{prop2}:} (a) Following the proof of Proposition \ref{prop1}, we obtain $\bar F_{X_i}(z)(q_i) = w_i/n + o(1/n)$ where $q_i = q_i(n) = (\zeta_i n/w_i)^{1/\nu_i}$. Next, choose $\epsilon^* > 0$ such that $(1+\epsilon^*)\xi_k < \nu_k$ and define
$$
G_k(y,z) = \left\{1 - F_{Z_k}\left(\frac{z}{y^{1/\nu_k}}\right)\right\}f_P(y).
$$
It follows that $\bar F_{X_k}(z) = \int_0^{\infty} G_k(y,z)\d y$ 
%$$\bar F_{X_k}(z) = \int_0^{\infty} G_k(y)\d y =  \int_0^{\infty}\left\{1 - F_{Z_k}\left(\frac{z}{y^{1/\nu_k}}\right)\right\}f_P(y) \d y,$$
and, since $z y^{-1/\nu_k} > z^{1 - (1+\epsilon^*)\xi_k/\nu_k}$ for $y < z^{\xi_k(1+\epsilon^*)}$, 
\begin{align*}
\int_0^{z^{\xi_k(1+\epsilon^*)}}G_k(y,z) \d y &= C_kz^{-\xi_k}\int_0^{z^{\xi_k(1+\epsilon^*)}} y^{\xi_k/\nu_k}f_P(y) \d y + o\left(z^{-\xi_k}\right) 
 = \zeta_k^* z^{-\xi_k} + o\left(z^{-\xi_k}\right),\\
\int_{z^{\xi_k(1+\epsilon^*)}}^{\infty}G_k(y,z) \d y &\leq C_P\int_{z^{\xi_k(1+\epsilon^*)}}^{\infty} \d y/y^2 = C_Pz^{-\xi_k(1+\epsilon^*)} = o(z^{-\xi_k}),
\end{align*}
where $\zeta_k^* = C_k\int_0^{\infty} y^{\xi_k/\nu_k}f_P(y) \d y$. Hence, if we let $q_k = q_k(n) = (\zeta_k^* n/w_k)^{1/\xi_k}$, then $\bar F_{X_k}(q_k) = w_k/n + o(1/n)$, and the stable tail dependence function of the pair $(X_i, X_k)^{\top}$ is
\begin{align*}
    \ell_{X_i,X_k}(w_i,w_k) &=   \lim_{n\to\infty} n\int_0^{\infty} \left\{1 - F_{Z_i,Z_k}\left(\frac{q_i}{y^{1/\nu_i}},\frac{q_k}{y^{1/\nu_k}} \right) \right\} f_P(y) \d y\\
    & = w_i + w_k - \lim_{n\to\infty} n \int_0^{\infty} G_{i,k}(y) \d y,
\end{align*}
where $G_{i,k}(y) = \bar F_{Z_i,Z_k}\left(\frac{q_i}{y^{1/\nu_i}},\frac{q_k}{y^{1/\nu_k}} \right)  f_P(y)$ and  $F_{Z_i, Z_k}$, $\bar F_{Z_i, Z_k}$ are the joint CDF and survival function of $(Z_i, Z_k)^{\top}$, respectively. Now let $0 < \alpha < 1$. We find:
\begin{align*}
    \lim_{n\to\infty} n\int_0^{n^{\alpha}}  G_{i,k}(y) \d y  & \leq \lim_{n\to\infty} n\int_0^{n^{\alpha}}  \bar F_{Z_i}\left(\frac{q_i}{y^{1/\nu_i}} \right)  f_P(y) \d y 
     \leq \left(\frac{w_i}{\zeta_i}\right)^{1+\epsilon}\lim_{n\to\infty} n^{-\epsilon}\int_0^{n^{\alpha}}  y^{1+\epsilon}  f_P(y) \d y,\\
     & \leq  C_P\left(\frac{w_i}{\zeta_i}\right)^{1+\epsilon} \lim_{n\to\infty} n^{-\epsilon}\int_0^{n^{\alpha}}  y^{\epsilon-1}  \d y = 0,\\
     \lim_{n\to\infty} n\int_{n^{\alpha}}^{n^{1+\epsilon^*}}  G_{i,k}(y) \d y & \leq \lim_{n\to\infty} n\int_{n^{\alpha}}^{n^{1+\epsilon^*}}  \bar F_{Z_k}\left(\frac{q_k}{y^{1/\nu_k}} \right)  f_P(y) \d y 
     \leq \frac{C_kw_k}{\zeta_k^*}\lim_{n\to\infty}\int_{n^{\alpha}}^{\infty}y^{\xi_k/\nu_k}f_P(y)\d y\\
     & \leq \frac{C_PC_kw_k}{\zeta_k^*}\lim_{n\to\infty}\int_{n^{\alpha}}^{\infty}y^{\xi_k/\nu_k-2}\d y = 0,\\
    \lim_{n\to\infty} n\int_{n^{1+\epsilon^*}}^{\infty}  G_{i,k}(y) \d y & \leq C_P\lim_{n\to\infty} n \int_{n^{1+\epsilon^*}}^{\infty}\d y/y^2 = 0.
\end{align*}
It implies that $\ell_{X_i,X_k}(w_i,w_k) = w_i + w_k$, so that $(X_i,X_k)^{\top}$ has no upper tail dependence.

\medskip
(b) From (a), we can take $q_i = q_i(n) = (\zeta_i^* n/w_i)^{1/\xi_i}$ and $q_k = q_k(n) = (\zeta_k^* n/w_k)^{1/\xi_k}$ and write
\begin{align*}
    \ell_{X_i,X_k}(w_i,w_k) &=   \lim_{n\to\infty} n\int_0^{\infty} \left\{1 - F_{Z_i,Z_k}\left(\frac{q_i}{y^{1/\nu_i}},\frac{q_k}{y^{1/\nu_k}} \right) \right\} f_P(y) \d y\\
    & = w_i + w_k - \lim_{n\to\infty} n \int_0^{\infty} G_{i,k}(y) \d y.
\end{align*}
Let $\bar C_{Z_i,Z_k}$ be the survival copula of $(Z_i, Z_k)$. We find:
\begin{align*}
    \lim_{n\to\infty}
    n\int_0^{n^{1+\epsilon^*}} G_{i,k}(y) \d y
     & = \lim_{n\to\infty} \int_0^{n^{1+\epsilon^*}}n\bar C_{Z_i,Z_k}\left(\frac{C_iw_i}{n\zeta_i^*}y^{\xi_i/\nu_i},\frac{C_kw_k}{n\zeta_k^*}y^{\xi_k/\nu_k}\right)f_P(y) \d y\\
    & \leq \int_0^{\infty} \lim_{n\to\infty} n \bar C_{Z_i,Z_k}\left(\frac{C_iw_i}{n\zeta_i^*}y^{\xi_i/\nu_i},\frac{C_kw_k}{n\zeta_k^*}y^{\xi_k/\nu_k}\right)f_P(y) \d y = 0,\\ \lim_{n\to\infty}n\int_{n^{1+\epsilon^*}}^{\infty}  G_{i,k}(y) \d y & \leq C_P\lim_{n\to\infty} n \int_{n^{1+\epsilon^*}}^{\infty}\d y/y^2 = 0,
\end{align*}
where the first inequality follows from the dominated convergence theorem since $$n\bar C_{Z_i,Z_k}\left(\frac{C_iw_i}{n\zeta_i^*}y^{\xi_i/\nu_i},\frac{C_kw_k}{n\zeta_k^*}y^{\xi_k/\nu_k}\right)f_P(y) \leq n\frac{C_iw_i}{n\zeta_i^*}y^{\xi_i/\nu_i}f_P(y) = \frac{C_iw_i}{\zeta_i^*}y^{\xi_i/\nu_i}f_P(y),$$  
and $\int_0^{\infty} y^{\xi_i/\nu_i}f_P(y) \d y < \infty$. 
Again, it implies that $\ell_{X_i,X_k}(w_i,w_k) = w_i + w_k$, so that $(X_i,X_k)^{\top}$ has no upper tail dependence. The proof is now complete. \hfill $\Box$

Proposition \ref{prop2} shows that the parameters $\nu_1, \ldots, \nu_d$, which control the tail behaviour of the Pareto factors $(P^{1/\nu_1}, \ldots, P^{1/\nu_d})^{\top}$, can be chosen such that only certain pairs of variables  $(X_i, X_k)^{\top}$  have tail dependence, while other pairs do not. This provides greater flexibility in modeling the upper tail of the joint distribution of $\XX$. In the next section, we illustrate this idea with an example of a model that is a special case of the Pareto-mixture model \eqref{main-eq}.

\emph{Remark 1:} Note that the lower-tail behavior of $\XX$ can be analyzed by examining the upper-tail behavior of $-\XX$ in the upper tail. Since $-X_i = (-Z_i)P^{1/\nu_i}$, the results of Propositions \ref{prop1} and \ref{prop2} can be applied directly to study the lower tail of $\XX$ by replacing $\ZZ$ with $-\ZZ$, provided that $\bar F_{Z_i}(0) < 1$ for all $i \in  \{1,\ldots, d\}$. 

\section{Special cases}
\label{sec-spec-case}

We now consider several interesting special cases of the model \eqref{main-eq} that result in skew-multivariate distributions of $\XX$.

\subsection{Grouped-$t$ distribution}

Consider the standard multivariate normal distribution $\ZZ \sim MVN(\boldsymbol{0}, \OO)$, where $\OO$ is the correlation matrix, and take $P = \sqrt{V}$ where $V$ follows the inverse gamma distribution: $V \sim \mathrm{Ig}(1/2,1/2)$. The density of $P$ is $f_P(z) = \sqrt{2/\pi} z^{-2}\exp(-0.5/z^2)$, and it satisfies Assumption \ref{ass1} and assumptions of Proposition \ref{prop1}. Note that $\Pr(\ZZ > \boldsymbol{0}) > 0$, and from Corollary \eqref{corol1}, the vector $\XX$ exhibits upper tail dependence, and its stable tail dependence function is given by
$$
\ell_{\XX}(w_1,\ldots,w_d) = \int_0^{\infty}\left[1 - \Phi_{\OO}\left\{\left(\frac{y}{w_1}\zeta_1\right)^{1/\nu_1}, \ldots, \left(\frac{y}{w_d}\zeta_d\right)^{1/\nu_d}\right\}\right] \d y\,,
$$
where $\Phi_{\OO}$ denotes the CDF of the standard multivariate normal distribution with the correlation matrix $\OO$, and $\zeta_i = 2^{\nu_i/2-1}\Gamma(\nu_i/2+1/2)/\sqrt{\pi}$, $i = 1,\ldots, d$. The copula of $\XX$ is the grouped-$t$
 copula \citep{Demarta.McNeil2005, Luo.Shevchenko2010}. Although the full likelihood for this model is not available in a simple form in the general case, approximate parameter estimates can be obtained using Kendall's measure of association. The respective limiting EV copula of $\XX$, $C_{\XX}^{\mathrm{EV}}$ is permutation asymmetric unless the parameters $\nu_1, \ldots, \nu_d$ are all equal. In this case, $C_{\XX}^{\mathrm{EV}}$ simplifies to the extremal-$t$ copula \citep{Nikoloulopoulos.Joe.ea2009}.

For a pair $(X_1,X_2)^{\top}$, we now compute the Pickands dependence function $A(t): [0,1] \to [1/2,1]$ \citep{Pickands1981}. The  function is convex and satisfies $\max(t, 1-t) \leq A(t) \leq 1$, and can be expressed in terms of the stable tail dependence function as $A(t) = \ell_{X_1,X_2}(1-t,t)$. This function $A(t)$ is symmetric about $t = 0.5$ when the EV copula $C_{X_1,X_2}^{\mathrm{EV}}$ is permutation symmetric.

\begin{figure}[ht]
    \centering
    \includegraphics[width=.5\linewidth]{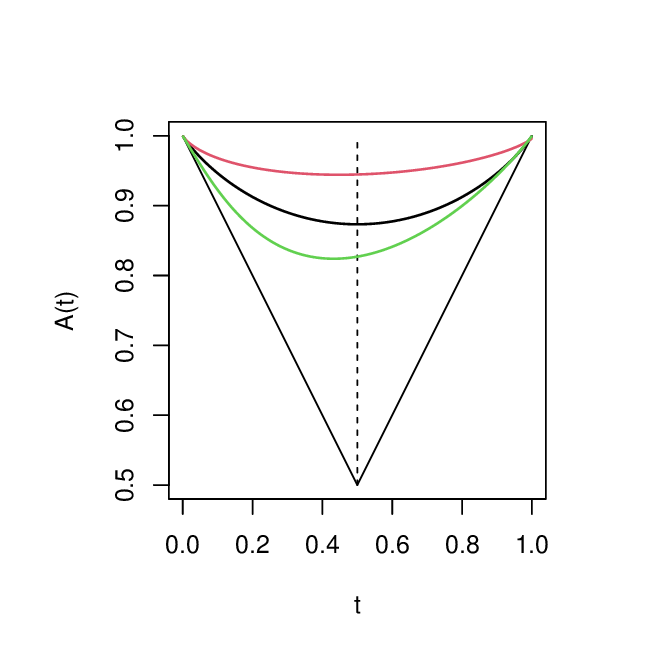}
    \hspace{-7mm}
    \includegraphics[width=.5\linewidth]{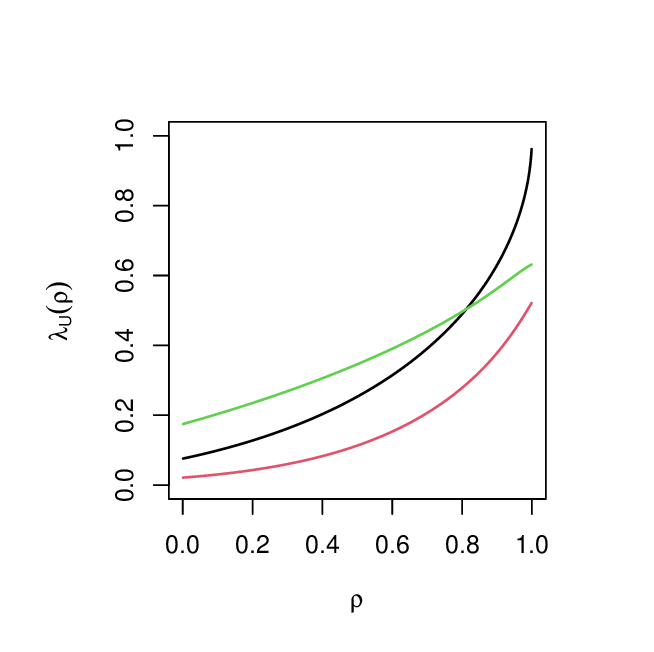}
    \vspace{-5mm}
    \caption{\footnotesize{Plots of 
    the Pickands dependence function $A(t)$ for $\rho = 0.5$ and $0 < t < 1$ (left), and the upper tail dependence coefficient $\lambda_U(\rho)$ for $0 < \rho <1$ (right), for the grouped-$t$ distribution with parameters $\Omega_{12} = \rho$ and three choice of degrees of freedom: $\nu_1 = \nu_2 = 4$ (black), $\nu_1 = 5, \nu_2 = 10$ (red), and $\nu_1 = 1, \nu_2 = 3$ (green).} }
    \label{fig-plot1}
\end{figure}

We also compute the upper tail dependence coefficient $\lambda_U$ for the pair $(X_1, X_2)^{\top}$. Figure \ref{fig-plot1} presents the plots of $A(t)$ for $0 < t < 1$ along with the corresponding value of $\lambda_U$ based on specific parameter values. The results demonstrate that the model can capture a broad range of upper tail dependence, with stronger dependence occurring when $\Omega_{1,2}$ is large and $\nu_1 = \nu_2$. An asymmetric Pickands function further shows that permutation asymmetry arises when $\nu_1 \neq \nu_2$. However, since $\ZZ \overset{d}{=} -\ZZ$, it follows that $\XX \overset{d}{=} -\XX$, implying that the model can represent only symmetric tail behavior.

%Figure \ref{fig-plot1} displays the plots of $\ell_{12}^U(w) := \ell_{X_1,X_2}(w,1) - w$ and $\ell_{21}^U(w) := \ell_{X_1,X_2}(1,w) - w$ for $1 \leq w \leq 10$, based on specific parameter values. The results show that the model is capable of capturing permutation asymmetry in the tails. \blue{However, because $\ZZ \overset{d}{=} -\ZZ$, it follows that $\XX \overset{d}{=} -\XX$, implying that the model can only represent symmetric tail behavior.}

\subsection{Skew-multivariate distribution with lower and upper tail dependence}

Again, consider the standard multivariate normal distribution $\ZZ^* \sim MVN(\boldsymbol{0}, \OO)$, where $\OO$ is the correlation matrix.  We define $\ZZ = \exp(\ZZ^* + \boldsymbol{\alpha}\ZM_{\tau})$, where $\boldsymbol{\alpha} = (\alpha_1, \ldots, \alpha_d)^{\top} \in \mathbb{R}^d$ is the vector of skewness parameters. Additionally, we let $P = \exp(\EE_U - \beta \EE_L)$, with $\beta > 0$, where $\EE_L, \EE_U$ are exponential $\mathrm{Exp}(1)$ random variables. The variable $\ZM_{\tau}$ follows a truncated normal distribution with the PDF $f_{\ZM}(z) = \phi(z)/\Phi(\tau)$ for $z > -\tau$, where $\tau \in \mathbb{R}$ is the truncation parameter.  Furthermore, we assume that $\ZZ^*, \EE_L, \EE_U, \ZM_{\tau}$ are independent. 

\begin{prop}
\label{prop3}
    %Consider the pair $(X_i,X_k)^{\top}$ for $i \neq k$, $i,k \in \{1,\ldots,d\}$. The pair $(X_i, X_k)^{\top}$ exhibits upper tail dependence and the corresponding tail dependence function is
    The stable tail dependence function of the vector $\XX$ is
    $$
    \ell_{\XX}(w_1,\ldots,w_d) = \sum_{l=1}^d w_l\Phi_{\OO_l^{\circ}}^{\mathrm{ESN}}\left(\frac{\bm \lambda_{l,-l}}{2} + \frac{1}{\bm \lambda_{l,-l}} \ln \frac{\tilde w_l}{\bm {\tilde w_{-l}}};{\bm \alpha_l^{\circ}},\tau_l^{\circ}\right), \quad {\tilde w_l = \frac{w_l}{\Phi(\tau + \alpha_l \nu_l)}}\,,  %\ell_{X_i, X_k}(w_i,w_k) = \sum_{l \in \{i,k\}} w_l \Phi_1\left(\frac{\lambda_{ik}}{2} + \frac{1}{\lambda_{ik}} \ln \frac{\tilde w_l}{\tilde w_{-l}};0,1,\alpha^*_l, \tau^*_l\right)\,,  \quad \tilde w_l = w_l/\Phi(\tau+\alpha_l \nu_l),
    $$
    where $\Phi_{\OO_l^{\circ}}^{\mathrm{ESN}}(\cdot;  {\bm\alpha_l^{\circ}}, {\tau_l^{\circ}})$ is a multivariate extended skew normal distribution function  with the shape-parameter vector ${\bm \alpha_l^{\circ}}$ and extension parameter $\tau_l^{\circ}$ \citep{Arellano.Genton2010}. Here  $\xx_{-l}$ denotes the vector $\xx$ with its $l$th component removed, $\OO_l^{\circ}$ is a $(d-1) \times (d-1)$ correlation matrix with entries
    $$(\OO_l^{\circ})_{ik} = \frac{\lambda_{il}^2 + \lambda_{kl}^2 - \lambda_{ik}^2}{2\lambda_{il}\lambda_{kl}}\,, \quad i, k \in \{1,\ldots, d\} \backslash l,$$ 
    and
    $$
    \quad {\bm \alpha^{\circ}_l} = -\frac{{\bm \omega_l(\OO_l^{\circ}})^{-1}}{\{1 - {\bm \omega_l(\OO_l^{\circ}})^{-1}{\bm \omega_l^{\top}}\}^{1/2}}\,, \quad \tau_l^{\circ} = \frac{\tau+\alpha_l \nu_l}{\{1 - {\bm \omega_l(\OO_l^{\circ}})^{-1}{\bm \omega_l^{\top}}\}^{1/2}}\,,
    $$
    with ${\bm \omega_l = (\alpha_l\nu_l - {\bm \alpha_{-l}}{\bm \nu_{-l}})}/{\bm \lambda_{l,-l}}$, $\lambda_{ik} = \{\Psi_{ik}^2 +(\alpha_{i}\nu_{i} - \alpha_k \nu_k)^2\}^{1/2}$, $\Psi_{ik} = (\nu_i^2 -2\Omega_{ik}\nu_i\nu_k + \nu_k^2)^{1/2}$ for $i \neq k$.
\end{prop}

The proof is in Appendix \ref{appx-prop3}. 

%Note that $\bar F_{Z_i}(z) = o(z^{-\nu})$ as $z \to \infty$ for any $\nu > 0$ and the density of $P$ is $f_P(z) = z^{1/\beta-1}/(\beta+1)$ for $z \leq 1$ and $f_P(z) = z^{-2}/(\beta+1)$ for $z > 1$, so that it satisfies Assumption \ref{ass1} and assumptions of Proposition \ref{prop1}. The vector $\XX$ exhibits upper tail dependence, and some calculations show that the EV limiting distribution of $\XX$ is the extremal extended skew-normal distribution \citep{Beranger.Padoan.ea2019}. For simplicity, consider the $(i,k)$-th marginal, where the corresponding tail dependence function is
    %$$
    %\ell_{X_i, X_k}(w_i,w_k) = \sum_{l \in \{i,k\}} w_l \Phi_1\left(\frac{\lambda_{ik}}{2} + \frac{1}{\lambda_{ik}} \ln \frac{\tilde w_l}{\tilde w_{-l}};0,1,\alpha^*_l, \tau^*_l\right)\,,  \quad \tilde w_l = w_l/\Phi(\tau+\alpha_l \nu_l),
    %$$
    %where $\Phi_1(\cdot; \mu, \sigma^2, \alpha, \tau)$ is the extended skew normal distribution function  with the shape parameter $\alpha$ and extension parameter $\tau$ \citep{Arellano.Genton2010}. Here $$\alpha^*_l = \frac{\alpha_{-l}\nu_{-l} - \alpha_l \nu_l}{\Psi_{ik}}, \quad \tau_l^* = (\tau+\alpha_l \nu_l)\left\{1+(\alpha^*_l)^2\right\}^{1/2}\,, \quad l \in \{i,k\},$$ where $\Psi_{ik} = (\nu_i^2 - 2\Omega_{i,k}\nu_i\nu_k + \nu_k^2)^{1/2}$, \ $\lambda_{ik} = \{\Psi_{ik}^2+(\alpha_{-l}\nu_{-l} - \alpha_l \nu_l)^2\}^{1/2}$. The proof is in Appendix. 
The limiting EV distribution of $\XX$ is the extremal extended skew-normal distribution \citep{Beranger.Padoan.ea2019}. Since $\vartheta_{\XX} = \ell_{\XX}(1, \ldots, 1) < d$, the vector $\XX$ exhibits upper tail dependence. The limiting EV distribution of $\XX$ simplifies to the H\"usler-Reiss distribution if $\boldsymbol{\alpha} = \boldsymbol{0}$, and the vector of skewness parameters $\boldsymbol{\alpha}$ controls permutation asymmetry in the general case.

\emph{Remark 2:} Consider $\nu_1 = \cdots = \nu_d = 1/\alpha_U$ and $\beta = \alpha_L/\alpha_U$, where $\alpha_L, \alpha_U > 0$ are given parameters. Define
\begin{equation}
     \label{eq-skewmodel1}
    \WW = \ln \XX = \ZZ^* + \boldsymbol{\alpha}\ZM_{\tau} + \alpha_U \EE_U - \alpha_L \EE_L.
\end{equation} 
The vector $\WW$ is a monotone transformation of $\XX$, so it has the same copula as $\XX$, $C_{\XX}$. This model can be viewed as an extension of the factor copula model for spatial data proposed by \cite{Krupskii.Huser.ea2018}, where the standard multivariate normal distribution $\ZZ^*$ replaced by $\ZZ^* + \boldsymbol{\alpha}\ZM_{\tau}$ which follows the multivariate extended skew-normal distribution \citep{Azzalini.Capitanio1999,Arellano.Genton2010}.  
The limiting behavior of $\WW$ in the upper tail is characterized by the extremal extended skew-normal distribution which depends on the vector of skewness parameters $\boldsymbol{\alpha}$ and the upper tail parameter $\alpha_U$. 
Furthermore, one can write $-\WW = - \ZZ^* - \boldsymbol{\alpha} \ZM_{\tau} + \alpha_L\EE_L - \alpha_U\EE_U$, so that the limiting behavior of $\WW$ in the lower tail is similarly governed by the extremal extended skew-normal distribution, but now with the vector of skewness parameters $-\boldsymbol{\alpha}$ and the lower tail parameter $\alpha_L$. Consequently, the distribution of $\WW$ provides greater flexibility both in the lower and upper tails.

\begin{figure}[h!]
    \centering
    \includegraphics[width=.5\linewidth]{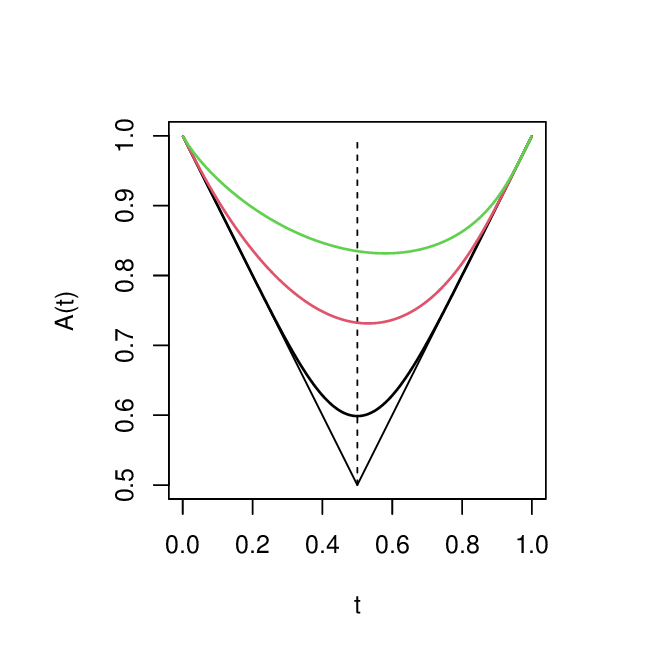}
    \hspace{-7mm}
    \includegraphics[width=.5\linewidth]{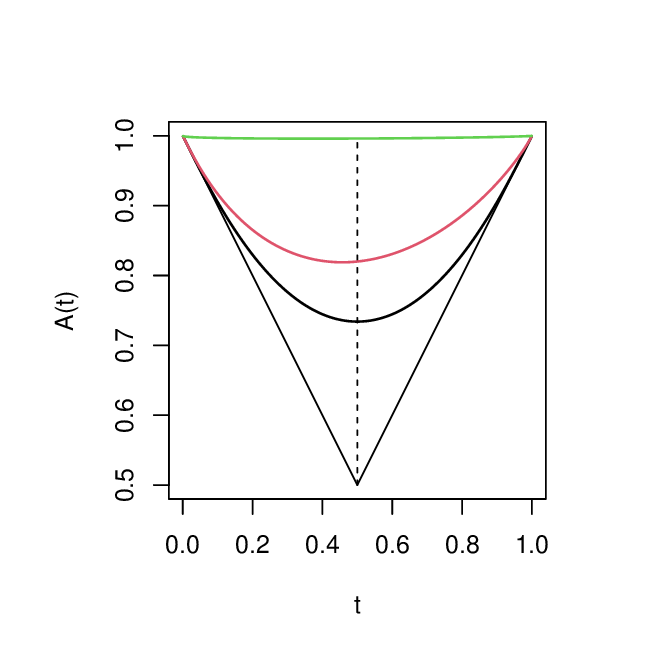}\vspace{-14mm}
    \includegraphics[width=.5\linewidth]{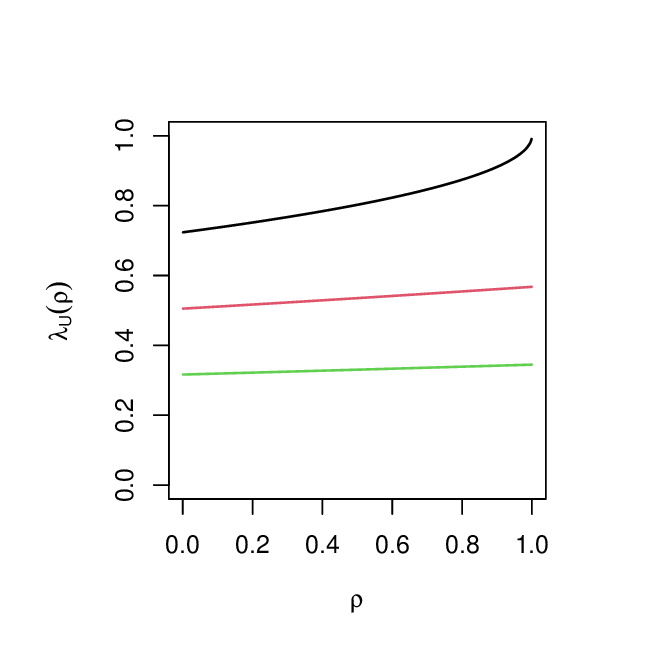}\hspace{-7mm}
    \includegraphics[width=.5\linewidth]{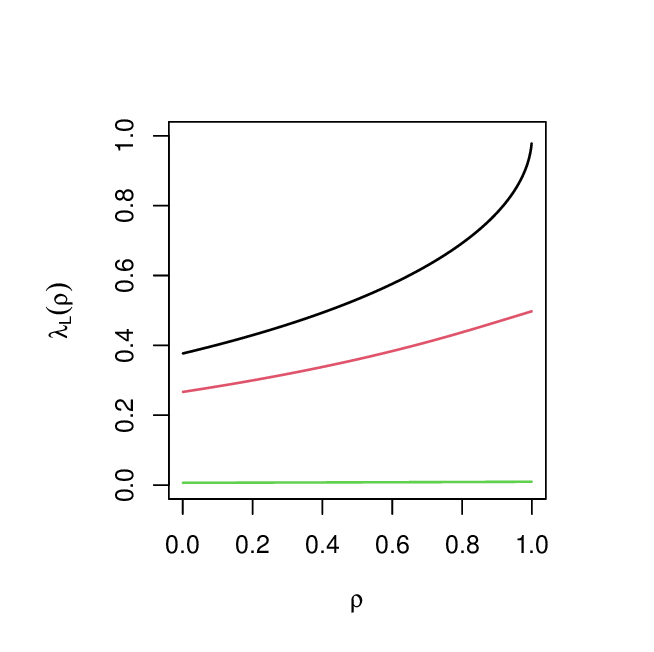}\vspace{-5mm}
    \caption{\footnotesize{Top row: Pickands dependence function $A(t)$ for $\rho = 0.5$ and $0 < t < 1$ for the pairs $(W_1,W_2)^{\top}$ (left) and $(-W_1,-W_2)^{\top}$. Bottom row: upper and lower tail dependence coefficients, $\lambda_U(\rho)$ and $\lambda_L(\rho)$ for $0 < \rho < 1$ (left and right, respectively) for model \eqref{eq-skewmodel1} with parameters $\Omega_{12} = \rho, \tau = 0, \alpha_U = 2, \alpha_U = 0.8$, and $\alpha_1 = \alpha_2 = 0$ (black), $\alpha_1 = 0, \alpha_2 = 3$ (red), and $\alpha_1 = -3, \alpha_2 = 3$ (green).}}
    \label{fig-plot2}
\end{figure}

Figure \ref{fig-plot2} presents the Pickands dependence function $A(t)$ for $0 < t < 1$  for the pairs $(W_1,W_2)^{\top}$ and $(-W_1, -W_2)^{\top}$, together with the corresponding  upper and lower tail dependence coefficients, $\lambda_U$ and $\lambda_L$, computed for selected parameter values. The results illustrate that the model can capture a wide range of both upper and lower tail dependence, with stronger dependence arising when $\Omega_{1,2}$ is large and $\alpha_1 = \alpha_2$. An asymmetric Pickands function further indicates that permutation asymmetry appears in both tails when $\alpha_1 \neq \alpha_2$. As the plots illustrate, this model can also represent asymmetry between the lower and upper tails --- an important feature not accommodated by the grouped-$t$ distribution. In Section \ref{subsec-mle-closed} we show that the copula density of $\WW$ as given in \eqref{eq-skewmodel1}, is available in a simple form, making likelihood-based inference straightforward for this model.

%of $\ell_{12}^U(w) := \ell_{X_1,X_2}(w,1) - w$ and $\ell_{21}^U(w) := \ell_{X_1,X_2}(1,w) - w$, as well as $\ell_{12}^L(w) := \ell_{-X_1,-X_2}(w,1) - w$ and $\ell_{21}^L(w) := \ell_{-X_1,-X_2}(1,w) - w$
%for $1 \leq w \leq 10$, based on specific parameter values. The results illustrate that the model can capture permutation asymmetry in both tails and can account for tail asymmetry as well, unlike the grouped-$t$ distribution. In Section \ref{subsec-mle-closed} we show that the copula density of $\WW$ as given in \eqref{eq-skewmodel1}, is available in a simple form, making likelihood-based inference straightforward for this model. 

\subsection{Skew-multivariate distribution with flexibility in the upper tail}

Consider the standard multivariate normal distribution $\ZZ^* \sim MVN(\boldsymbol{0}, \OO)$, where $\OO$ is the correlation matrix.  We define $\ZZ = \exp(\ZZ^* + \boldsymbol{\alpha}\EE_0)$, where $\boldsymbol{\alpha} > 0$ and $\EE_0$ is an exponential $\mathrm{Exp}(1)$ random variable. Additionally, we assume that $P$ follows the Pareto distribution with the PDF $f_P(z) = z^{-2}$ for $z > 1$.  Furthermore, we assume that $\ZZ^*$ and $ \EE_0$ are independent. 

\begin{prop}
    \label{prop4a}
    Consider the pair $(X_i, X_k)^{\top}$ for $i \neq k$, $i,k \in \{1,\ldots, d\}$. If $\alpha_i \nu_i < 1$ and $\alpha_k\nu_k > 1$ or if $\alpha_i \nu_i > 1$ and $\alpha_k\nu_k < 1$, then $(X_i,X_k)^{\top}$ has no upper tail dependence. On the other hand, if $\alpha_i\nu_i < 1$  and $\alpha_k\nu_k < 1$ or $\alpha_i\nu_i > 1$  and $\alpha_k\nu_k > 1$, then $(X_i, X_k)^{\top}$ has upper tail dependence. In particular, if $\alpha_i\nu_i < \alpha_k\nu_k < 1$, then
    \begin{multline*}
\ell_{X_i, X_k}(w_i,w_k) =w_i\Phi\left(\varrho_i\right) + w_k\Phi\left(\varrho_k\right) + w_k \left(\frac{\tilde w_k}{\tilde w_i}\right)^{1/(\xi_k/\xi_i-1)}\exp\left\{\frac{0.5\Psi_{ik}^2\xi_k/\xi_i}{(\xi_k/\xi_i-1)^2}\right\}\Phi\left(-\varrho_k-\frac{\Psi_{ik}}{\xi_k/\xi_i-1}\right)\\
- w_i \left(\frac{\tilde w_i}{\tilde w_k}\right)^{1/(\xi_i/\xi_k-1)}\exp\left\{\frac{0.5\Psi_{ik}^2\xi_i/\xi_k}{(\xi_i/\xi_k-1)^2}\right\}\Phi\left(\varrho_i+\frac{\Psi_{ik}}{\xi_i/\xi_k-1}\right)
\,,
\end{multline*}
where $\Psi_{ik}$ is defined in the previous example, $\varrho_l = \frac{\Psi_{ik}}{2} + \frac{1}{\Psi_{ik}}\ln \frac{\tilde w_{l}}{\tilde w_{-l}}$, $\xi_l  = 1/(1 - \alpha_l \nu_l)$, and $\tilde w_l = w_l/\xi_l$. Here, $-l = \{i,k\} \backslash l$ and $\l \in \{i,k\}$. A similar result holds for $\alpha_k\nu_k \leq \alpha_i\nu_i < 1$. If $\alpha_l\nu_l > 1$ for $l \in \{i,k\}$, the above formula can be used to compute $\ell_{X_i,X_k}(w_i,w_k)$ by replacing $\alpha_l$ with $1/\nu_l$ and $\nu_l$ with $1/\alpha_l$.
\end{prop}

The proof is in Appendix \ref{appx-prop4a}. 
The parameters $\boldsymbol{\alpha}$ and $\boldsymbol{\nu}$ control all pairs of variables $(X_i, X_k)^{\top}$ with upper tail dependence which provides greater flexibility in the upper tail.  Consequently, the limiting EV distribution of $(X_i, X_k)^{\top}$ is permutation asymmetric, unless $\alpha_k\nu_k = \alpha_i \nu_i$, in which case it simplifies to the H\"usler-Reiss distribution.

\emph{Remark 3:} Consider $\nu_i  = 1/\beta_i$, where $\beta_i > 0$, $i = 1, \ldots, d$ are given parameters. Define
\begin{equation}
    \label{eq-skewmodel2}
    \WW = \ln \XX = \ZZ^* + \boldsymbol{\alpha} \EE_0 + \boldsymbol{\beta} \EE_1,
\end{equation}
where $\EE_1 = \ln P$ follows an exponential $\mathrm{Exp}(1)$ distribution. Again, the vector $\WW$ has the same copula as $\XX$, $C_{\XX}$. The limiting behavior of $\WW$ in the upper tail is characterized by the vectors $\boldsymbol{\alpha}$ and $\boldsymbol{\beta}$. The next Proposition shows that $\WW$ exhibits no lower tail dependence, meaning this model is not suitable for data that exhibit both lower and upper tail dependence.

\begin{prop}
\label{prop4b}
For any $i \neq  k$, $i, k \in \{1,\ldots, d\}$, the lower tail dependence coefficient of the pair $(W_i, W_k)^{\top}$ is zero.
\end{prop}
The proof of Proposition \ref{prop4b} is provided in Appendix \ref{appx-prop4b}.

Figure \ref{fig-plot3} displays the Pickands dependence function $A(t)$ for $0 < t < 1$ for the pair $(W_1, W_2)^{\top}$, along with the corresponding upper tail dependence coefficient $\lambda_U$, computed for selected parameter values. The results show that the model can capture a wide range of upper tail dependence, with stronger dependence occurring when $\Omega_{1,2}$ is large and $\alpha_1 = \alpha_2$ and $\beta_1 = \beta_2$. An asymmetric Pickands function further indicates that pronounced permutation asymmetry emerges when $\beta_1 \neq \beta_2$. In Section \ref{subsec-mle-closed} we show that the copula density of $\WW$ as defined in \eqref{eq-skewmodel2}, is available in a simple form, similar to the model in \eqref{eq-skewmodel1}, which facilitates straightforward likelihood-based inference for this model.

\begin{figure}[ht]
    \centering
    \includegraphics[width=.5\linewidth]{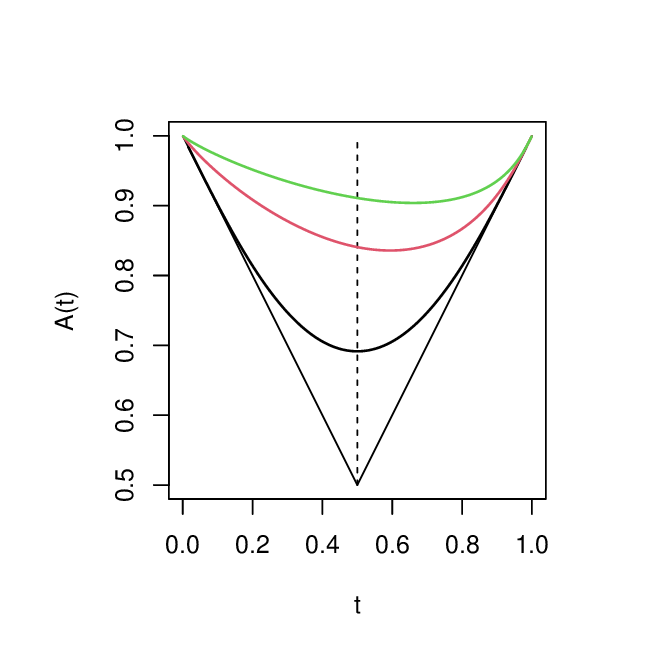}
    \hspace{-7mm}
    \includegraphics[width=.5\linewidth]{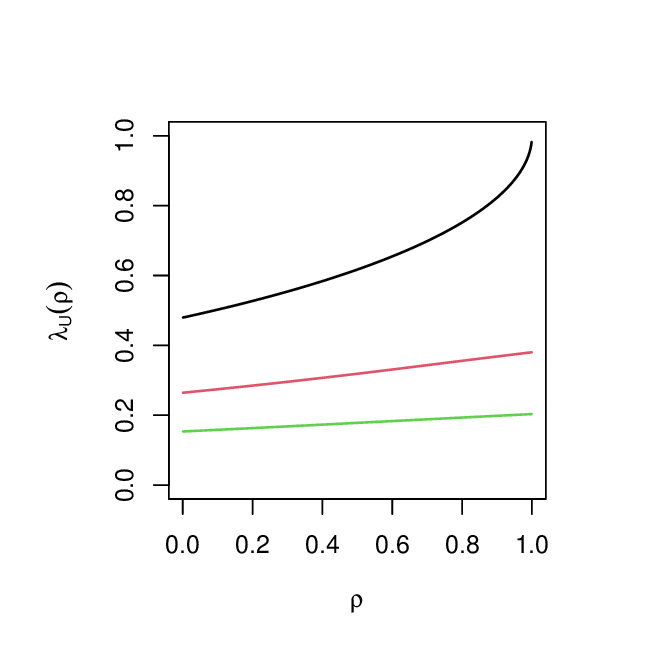}\vspace{-5mm}
    \caption{\footnotesize{\footnotesize{Plots of the Pickands dependence function for $\rho = 0.5$ and $0 <  t < 1$ (left) and the upper tail dependence coefficient $\lambda_U(\rho)$ for $0 < \rho < 1$ (right), for model \eqref{eq-skewmodel2} with  parameters $\Omega_{12} = \rho, \ \alpha_1 = \alpha_2 = 1$, and $\beta_1 = \beta_2 = 0.3$ (black), $\beta_1 = 0.3, \beta_2 = 0.9$ (red), and $\beta_1 = 0.1, \beta_2 = 0.95$ (green).}} }
    \label{fig-plot3}
\end{figure}

\section{Inference}
\label{sec-infer}

\subsection{Likelihood function}
\label{sec-mle}

Let $\{(x_{i1},\ldots,x_{id})^{\top}\}_{i=1}^n$ be a sample from a multivariate CDF $G_{\XX}$ that shares the same copula as $\XX$ defined in \eqref{eq-copmodel}, $C_{\XX}$, and let  $g_{\XX}$ be the respective multivariate PDF. Note that  
the marginal CDFs, $G_1, \ldots, G_d$, do not have to be the same as the marginal CDFs of $\XX$, thus providing greater flexibility of the model. We can write $$g_{\XX}(x_1, \ldots, x_d; \tht) = c_{\XX}\left\{G_1(x_1; \tht_M), \ldots, G_d(x_d; \tht_M); \tht_C\right\} \prod_{j=1}^d g_j (x_j; \tht_M), \quad \tht = (\tht_M^{\top}, \tht_C^{\top})^{\top},$$
where $\tht_M$ and $\tht_C$ are the vectors of marginal and copula parameters, respectively. From \eqref{eq-stdf}, the log-likelihood for this model is
\begin{align}    
    L(\tht) = \sum_{i=1}^n \ln g_{\XX}(x_{i1}, \ldots, x_{id};\tht) =& \sum_{i=1}^n \ln f_{\XX}\left[F_{X_1}^{-1}\left\{G_1(x_{i1}; \tht_M); \tht_C\right\}, \ldots, F_{X_d}^{-1}\left\{G_d(x_{id}; \tht_M);\tht_C\right\}; \tht_C\right] \nonumber\\
    &- \sum_{i=1}^n\sum_{j=1}^d \ln f_{X_j}\left[F_{X_j}^{-1}\left\{G_j(x_{ij};\tht_M);\tht_C\right\}\right] + \sum_{i=1}^n\sum_{j=1}^d \ln g_j(x_{ij};\tht_M). \label{eq-loglik}
\end{align}

Under regularity conditions \citep{White1982, Vuong1982}, the maximum likelihood estimator is consistent and asymptotically normal provided that the copula $C_{\XX}$ and marginal CDFs, $G_1, \ldots, G_d$, are correctly specified. Alternatively, marginal parameters $\tht_M$ can be estimated by maximizing the marginal log-likelihood:
$$
L_M(\tht_M) = \sum_{i=1}^n\sum_{j=1}^d \ln g_j(x_{ij};\tht_M).
$$
The copula parameters $\tht_C$ can then be estimated by maximizing the copula log-likelihood using the estimates of $\tht_M$, $\widehat\tht_M$, obtained in the first step:  
\begin{equation}
\label{eq-coplik}
L_C(\tht_C) =  \sum_{i=1}^n \ln f_{\XX}\left[F_{X_1}^{-1}(u_{i1};\tht_C), \ldots, F_{X_d}^{-1}(u_{id};\tht_C); \tht_C\right\} - \sum_{i=1}^n\sum_{j=1}^d \ln f_{X_j}\left[F_{X_j}^{-1}(u_{ij};\tht_C)\right],    
\end{equation}
where $u_{ij} = G_j(x_{ij}; \widehat\tht_M)$, $j \in \{1,\ldots, d\}$. The two-step approach, called the Inference Method for Margins (IFM), is more computationally tractable since the model parameters are estimated in two steps, and the IFM estimator in consistent and asymptotically normal \citep{Joe.Xu1996}.

Finally, the marginal distributions can be estimated nonparametrically and used to transform the data to the uniform 
$(0,1)$ scale. A common approach is to apply component-wise ranks:$$
u_{ij} = \frac{\mathrm{rank}(x_{ij}) - 0.5}{n}, \quad i = 1,\ldots, n.
$$
The copula log-likelihood $L_C(\tht_C)$ in \eqref{eq-coplik}  can then be maximized to estimate the copula parameters. This semiparametric method avoids specifying models for the marginal distributions and is computationally efficient. Its asymptotic properties have been investigated in \cite{Genest.Ghoudi.ea1995,Shih.Louis1995}. 

%For further details on the asymptotic properties of these alternative estimators, refer to Sections  5.4, 5.5 and 5.9 of \cite{Joe2015}.

\subsection{Models with tractable likelihoods}
\label{subsec-mle-closed}

From \eqref{eq-loglik} and \eqref{eq-coplik}, it is evident that a computationally tractable likelihood requires the joint PDF $f_{\XX}$ and marginal CDFs $F_{X_1}, \ldots, F_{X_d}$ to be available in a simple form, while the inverse CDFs $F_{X_1}^{-1}, \ldots, F_{X_d}^{-1}$ can be computed numerically. We will now provide explicit expressions for these functions for the models in \eqref{eq-skewmodel1} and \eqref{eq-skewmodel2}, where we replace $\XX$ with $\WW = \ln \XX$, as $\XX$ and $\WW$ share the same copula.

\subsubsection*{Skew-multivariate distribution with lower and upper tail dependence}

First, we derive the CDF of $W_i$ as defined in \eqref{eq-skewmodel1}. Note that $\EE^* = \alpha_U\EE_U - \alpha_L\EE_L$ follows an asymmetric Laplace distribution with the PDF $f_{\EE^*}(z) = \exp(z/\alpha_L)/(\alpha_L+\alpha_U)$ for $z \leq 0$ and $f_{\EE^*}(z) = \exp(-z/\alpha_U)/(\alpha_L+\alpha_U)$ for $z > 0$. Additionally, $V_i = Z^* + \alpha_i \ZZ_{\tau}$ follows an extended skew-normal distribution with the CDF $F_{V_i}(z) = \Phi_{-\alpha_i/\tilde\alpha_i}\left(z/\tilde\alpha_i, -\tau\right)/\Phi(-\tau)$, where $\tilde\alpha_i = (1+\alpha_i^2)^{1/2}$ and $\Phi_{\rho}$ denotes the CDF of a standard bivariate normal distribution with the correlation $\rho$. Thus, it follows that 
$$
    F_{W_i}(z) = C^*\int_{0}^{\infty}\Phi_{-\frac{\alpha_i}{\tilde\alpha_i}}\left(\frac{z-v}{\tilde\alpha_i}, -\tau\right) \exp\left(-\frac{v}{\alpha_U}\right)\d v + C^*\int_{-\infty}^0\Phi_{-\frac{\alpha_i}{\tilde\alpha_i}}\left(\frac{z-v}{\tilde\alpha_i}, -\tau\right) \exp\left(\frac{v}{\alpha_L}\right)\d v \,,
$$
where $C^* = 1/\{(\alpha_L+\alpha_U)\Phi(-\tau)\}$. Using the integration by parts formula, we find
\begin{multline*}
F_{W_i}(z) = C^*(\alpha_L+\alpha_U)\Phi_{-\frac{\alpha_i}{\tilde\alpha_i}}\left(\frac{z}{\tilde\alpha_i}, -\tau\right) - C^*\alpha_U\exp\left(\frac{\alpha_i^2+1}{2\alpha_U^2} - \frac{z}{\alpha_U}\right)\Phi_{-\frac{\alpha_i}{\tilde\alpha_i}}\left(\frac{z}{\tilde\alpha_i} - \frac{\tilde\alpha_i}{\alpha_U}, -\tau+\frac{\alpha_i}{\alpha_U}\right)\\
    + C^*\alpha_L\exp\left(\frac{\alpha_i^2+1}{2\alpha_L^2} + \frac{z}{\alpha_L}\right)\Phi_{\frac{\alpha_i}{\tilde\alpha_i}}\left(-\frac{z}{\tilde\alpha_i} - \frac{\tilde\alpha_i}{\alpha_L}, -\tau-\frac{\alpha_i}{\alpha_L}\right)\,.
\end{multline*}

Similarly, one can derive the joint PDF of $\WW$:
\begin{align*}
    f_{\WW}(\zz) =& \frac{1}{\Phi(-\tau)}\int_{\tau}^{\infty}\int_0^{\infty}\phi_{\OO}(\zz - \boldsymbol{\alpha} w - v)\phi(w)\exp\left(-\frac{v}{\alpha_U}\right) \d v \d w\\
    &+ \frac{1}{\Phi(-\tau)}\int_{\tau}^{\infty}\int_{-\infty}^0\phi_{\OO}(\zz - \boldsymbol{\alpha} w - v)\phi(w)\exp\left(\frac{v}{\alpha_L}\right) \d v \d w\\
    =& \zeta^*\exp\left(\frac{Ax_1}{2}+\frac{By_1}{2}-\frac{x_1}{2\alpha_U}\right)\Phi_{\rho}\left\{x_1C_1,(\tau+y_1)D_1\right\} \\
     &+ \zeta^*\exp\left(\frac{Ax_2}{2}+\frac{By_2}{2}+\frac{x_2}{2\alpha_L}\right)\Phi_{-\rho}\left\{-x_2C_1,(\tau+y_2)D_1\right\}\,,
\end{align*}
where $\rho = -E\{C(D+1)\}^{-1/2}$, $C_1 = \{C(1-\rho^2)\}^{1/2}, D_1 = \{(D+1)(1-\rho^2)\}^{1/2}$ and
$$
A = \ii^{\top}\OO^{-1}\zz, \quad B = \zz^{\top}\OO^{-1}\aa, \quad C = \mathrm{tr}(\OO^{-1}), \quad D = \aa^{\top}\OO^{-1}\aa, \quad E = \ii^{\top}\OO^{-1}\aa,
$$
$$
x_1 = \frac{BE - A(D+1) + (D+1)/\alpha_U}{E^2 - C(D+1)}, \quad 
  x_2 = \frac{BE - A(D+1) - (D+1)/\alpha_L}{E^2 - C(D+1)}\,,
$$
$$
y_1 = \frac{AE - BC - E/\alpha_U}{E^2 - C(D+1)}\,, \quad
  y_2 = \frac{AE - BC + E/\alpha_L}{E^2 - C(D+1)}\,,
$$
$$
\zeta^* = C^*(2\pi)^{1/2-d/2}\{C(D+1)(1-\rho^2)\}^{-1/2}|\OO|^{-1/2}\exp(-.5\zz^{\top}\OO^{-1}\zz).
$$

\subsubsection*{Skew-multivariate distribution with flexibility in the upper tail}

From \cite{Mondal.Krupskiy.ea2024}, the marginal CDF of $W_i$ as defined in \eqref{eq-skewmodel2} is
$$
F_{W_i}(z)  = \Phi(z)  - \frac{\alpha_{i} \exp \left( -\frac{z}{\alpha_i} + \frac{0.5}{\alpha_i^2}\right)  \Phi\left(z - \frac{1}{\alpha_i}\right) - \beta_i \exp \left( -\frac{z}{\beta_i} + \frac{0.5}{\beta_i^2} \right) \Phi\left(z - \frac{1}{\beta_i}\right) }{\alpha_{i} - \beta_i}\,,
$$
and the joint PDF of $\WW$ is
\begin{align*}
    f_{\WW}(\zz) =& \int_{0}^{\infty}\int_0^{\infty}\phi_{\OO}(\zz - \boldsymbol{\alpha} v_1 - \boldsymbol{\beta} v_2)\exp\left(-v_1-v_2\right) \d v_1 \d v_2\\
    =& \zeta^*\Phi_{\rho}\left[x_0\{A(1-\rho^2)\}^{1/2}, y_0\{B(1-\rho^2)\}^{1/2}\right], 
\end{align*}
where $\rho = -C(AB)^{-1/2}$ and 
$$x_0 = \frac{B(z_a-1)-C(z_b-1)}{AB-C^2}\,, \ \   y_0 = \frac{-C(z_a-1)+A(z_b-1)}{AB-C^2}, \quad z_a = \zz^{\top}\OO^{-1}\boldsymbol{\alpha}, \ \  z_b = \zz^{\top}\OO^{-1}\boldsymbol{\beta},$$
$$
A = \boldsymbol{\alpha}^{\top}\OO^{-1}\boldsymbol{\alpha}, \quad B = \boldsymbol{\beta}^{\top}\OO^{-1}\boldsymbol{\beta}, \quad C = \boldsymbol{\alpha}^{\top}\OO^{-1}\boldsymbol{\beta},
$$
$$
\zeta^* = (2\pi)^{1-d/2}\{AB(1-\rho^2)\}^{-1/2}|\OO|^{-1/2}\exp\left\{0.5x_0(z_a-1) + 0.5y_0(z_b-1) - 0.5\zz^{\top}\OO^{-1}\zz\right\}.
$$
Note that, unlike the model considered in \cite{Mondal.Krupskiy.ea2024}, the joint density of 
$\WW$ in \eqref{eq-skewmodel2} involves only a bivariate normal CDF, which makes the likelihood approach computationally more tractable for this model. 

\subsubsection*{Improving stability of the code}

The marginal CDFs $F_{W_1}, \ldots, F_{W_d}$ and the joint PDF $f_{\WW}$ for the two models in \eqref{eq-skewmodel1} and \eqref{eq-skewmodel2} contain exponential terms, which may become very large for certain  parameter values (e.g., when $\alpha_L$ or $\alpha_U$ is very small, or when $\alpha_j$ or $\beta_j$ is very close to zero for some $j \in \{1, \ldots, d\}$). To prevent overflow problems in model \eqref{eq-skewmodel2}, the exponential term and the normal CDF can be combined as follows:
$$
\exp(x)\Phi(y) = \exp\left\{x + \ln \Phi(y)\right\},
$$
where $x = -z/\alpha_j + 0.5/\alpha_j^2$, $y = z - 1/\alpha_j$ or $x = -z/\beta_j + 0.5/\beta_j^2$, $y = z - 1/\beta_j$. In this cases, $x + \ln \Phi(y)$ is typically much smaller in magnitude than $x$ for very large $x$. The log-normal CDF $\ln \Phi(y)$ can be computed with very high precision in many statistical software packages.

In model \eqref{eq-skewmodel1}, the following approximation can be used:
$$
\frac{\Phi_{\rho}(x,y)}{\phi(y)} \approx \frac{\Phi(y)}{\phi(y)}\Phi\left\{\frac{x-\rho y}{(1-\rho^2)^{1/2}}\right\} + \rho \varepsilon(x,y;\rho)\exp\left[\ln \phi(x) + \ln \Phi\left\{\mu\right\} - \ln \phi(y) - \ln(-y)\right],
$$
where $\mu = \mu(x,y;\rho) = (y-\rho x)/(1-\rho^2)^{1/2}$ and
$$
\epsilon(x,y;\rho) = -y\frac{\Phi(y)}{\phi(y)} + y\left\{y\frac{\Phi(y)}{\phi(y)} + 1\right\}\exp\left\{\ln \phi(\mu) - \ln \Phi(\mu) + \mu\right\}(1-\rho^2)^{1/2}\,.
$$
This approximation performs very well for large negative values of $y$ \citep{Au2024}. Specifically, it can be applied to compute the marginal CDFs with  $x = -\tau + \alpha_i/\alpha_U$, $y = z/\tilde\alpha_i - \tilde\alpha_i/\alpha_U$, $\rho = -\alpha_i/\tilde\alpha_i$ and with $x = -\tau - \alpha_i/\alpha_L$, $y = -z/\tilde\alpha_i - \tilde\alpha_i/\alpha_L$, $\rho = \alpha_i/\tilde\alpha_i$.

\section{Simulation studies}
\label{sec-sim}

In this section, we examine the finite sample performance of the maximum likelihood estimators for the models \eqref{eq-skewmodel1} and \eqref{eq-skewmodel2},  which have tractable likelihoods (referred to as models M1 and M2, respectively).
For each model, we randomly select $d = 10, 20, 30, 50$ locations in the unit square $(0,1)^2$ and simulate $n=200$ samples of size $N=100, 500$. Let $\ss_i = (s_{i1}, s_{i2})^{\top}$ denote the location of the $i$th variable, $i = 1, \ldots, d$. We use the powered-exponential covariance function $\rho(h) = \exp\{-(h/\mu_1)^{\mu_2}\}$, $\mu_1 > 0, \mu_2 \in (0,2]$ for the correlation matrix $\OO$. The following parameters are used for the two models:
\begin{itemize}
    \item[M1] $\OO_{i,k} = \exp\left\{-\left(||\ss_i - \ss_k||/\mu_1\right)^{\mu_2}\right\}$ with $\mu_1 = 0.5, \mu_2 = 1.5$, $\alpha_L = 3, \alpha_U = 2, \boldsymbol{\alpha}_i = \alpha_0 + \alpha_1 s_{i1} + \alpha_2 s_{i2}$ with $\alpha_0 = 5$, $\alpha_1 = \alpha_2 = -2.5$. Here we consider $\tau = 0$; this parameter is not estimated.
    \item[M2] $\OO_{i,k} = \exp\left\{-\left(||\ss_i - \ss_k||/\mu_1\right)^{\mu_2}\right\}$ with $\mu_1 = 0.5, \mu_2 = 1.5$, $\boldsymbol{\alpha} = \alpha_0$ with $\alpha_0  = 2$,  $\boldsymbol{\beta} = \exp(\beta_0 + \beta_1s_{i1} + \beta_2s_{i2})$ with $\beta_0 = 1.5, \beta_1 = \beta_2 = -2.5$. 
\end{itemize}

For both models, we use the skew-$t$ marginal distribution \citep{Azzalini.Capitanio2003}, with $\nu_M = 5$ degrees of freedom, location parameter $\xi_M = 5$, scale parameter $\omega_M = 1$ and slant parameter $\alpha_M = 5$. To estimate the copula parameters, we use the IFM and semiparametric estimators discussed in Section~\ref{sec-mle}, along with the approximation to the bivariate normal probability outlined in Section~\ref{subsec-mle-closed} which helps enhance the stability of the estimation procedure.

\begin{table}[]
    \centering
    \begin{tabular}{c|ccccccc}
     & $\theta_1$ & $\theta_2$ & $\alpha_L$ & $\alpha_U$ & $\alpha_0$ & $\alpha_1$ & $\alpha_2$ \\
     \hline
     $d = 10$ & 1.16/0.13 &  0.12/0.04 
 & 1.39/0.38 & 0.97/0.34 & 4.09/1.51 & 1.87/0.78 & 1.85/0.75\\ 
     $d = 20$ & 0.55/0.04 &  0.04/0.01 
 & 1.11/0.30 & 0.76/0.27 & 3.08/1.44 & 1.35/0.59 & 1.39/0.59\\ 
     $d = 30$ & 0.59/0.13 &  0.03/0.01 
 & 1.18/0.35 & 0.83/0.33 & 4.00/0.99 & 1.71/0.45 & 1.72/0.46\\ 
     $d = 50$ & 0.54/0.03 &  0.02/0.01 
 & 1.22/0.29 & 0.90/0.29 & 3.53/0.26 & 1.53/0.20 & 1.54/0.20 \\
 \hline
     $d = 10$ & 2.44/0.86 &  0.30/0.13 
 & 1.87/1.02 & 1.30/0.75 & 4.59/3.23 & 2.14/1.49 & 2.08/1.51\\ 
     $d = 20$ & 2.01/0.76 &  0.28/0.11 
 & 2.01/1.31 & 1.49/1.00 & 4.91/3.68 & 2.26/1.60 & 2.28/1.60\\ 
     $d = 30$ & 1.95/0.87 &  0.27/0.11 
 & 2.12/1.63 & 1.55/1.16 & 5.56/4.66 & 2.51/1.98 & 2.51/1.94\\ 
     $d = 50$ & 2.17/1.10 &  0.27/0.10 
 & 2.30/2.05 & 1.64/1.42 & 5.69/5.28 & 2.55/2.28 & 2.54/2.30  
    \end{tabular}
    \caption{RMSEs of the copula parameter estimates for Model M1 obtained using the maximum likelihood methods described in Section \ref{sec-mle}: the IFM estimator (top) and the semiparametric estimator (bottom). The true values are $\mu_1 = 0.5, \mu_2 = 1.5$, $\alpha_L = 3$, $\alpha_U = 2$, $\alpha_0 = 5$, $\alpha_1 = \alpha_2 = -2.5$. The results are based on 200 simulated data sets with sample size $N = 100$/$N = 500$ and $d=10, 20, 30, 50$ randomly selected locations in $(0,1)^2$.}
    \label{tab1}
\end{table}

\begin{table}[]
    \centering
    \begin{tabular}{c|cccccc}
     & $\theta_1$ & $\theta_2$ & $\alpha_0$ & $\beta_0$ & $\beta_1$ & $\beta_2$ \\
     \hline
     $d = 10$ & 0.10/0.04 &  0.07/0.03 
 & 0.40/0.16 & 0.47/0.18 & 0.71/0.31 & 0.74/0.28 \\
     $d = 20$ & 0.08/0.03 &  0.03/0.02 
 & 0.39/0.17 & 0.35/0.11 & 0.45/0.20 & 0.51/0.19 \\
     $d = 30$ & 0.07/0.03 &  0.02/0.01 
 & 0.36/0.16 & 0.25/0.09 & 0.41/0.17 & 0.39/0.17 \\
     $d = 50$ & 0.07/0.03 &  0.02/0.01 
 & 0.37/0.17 & 0.22/0.09 & 0.29/0.14 & 0.30/0.16 \\
     \hline
     $d = 10$ & 0.29/0.07 &  0.19/0.07
 & 0.62/0.25 & 0.74/0.25 & 1.04/0.47 & 0.99/0.47 \\
     $d = 20$ & 0.26/0.07 &  0.18/0.07 
 & 0.65/0.28 & 0.57/0.27 & 0.83/0.46 & 0.85/0.44 \\
     $d = 30$ & 0.27/0.08 &  0.19/0.07 
 & 0.71/0.32 & 0.65/0.27 & 0.86/0.45 & 0.85/0.45 \\
     $d = 50$ & 0.28/0.10 &  0.20/0.07 
 & 0.75/0.36 & 0.75/0.29 & 1.03/0.40 & 1.05/0.40 
    \end{tabular}
    \caption{RMSEs of the copula parameter estimates for Model M2 obtained using the maximum likelihood methods described in Section \ref{sec-mle}: the IFM estimator (top) and the semiparametric estimator (bottom). The true values are $\mu_1 = 0.5, \mu_2 = 1.5$, $\alpha_0 = 2$, $\beta_0 = 1.5$, $\beta_1 = \beta_2 = -2.5$. The results are based on 200 simulated data sets with sample size $N = 100$/$N = 500$ and $d=10, 20, 30, 50$ randomly selected locations in $(0,1)^2$.}
    \label{tab2}
\end{table}

Tables \ref{tab1} and \ref{tab2} present the root mean squared errors (RMSEs) of the copula parameter estimates for the two models; we do not report results for the marginal parameters for the IFM method. For model M1, the estimates are less accurate when $N=100$ but improve substantially when $N=500$. Increasing the number of spatial locations also enhances the accuracy of the IFM estimates, especially for parameters $\theta_1$ and  $\theta_2$. For model M2, both the IFM and semiparametric estimators exhibit higher accuracy, and IFM estimates improve with both a larger sample sample size, $N$, and a greater number of locations, $d$. For the both models, however, the semiparametric estimates remain less accurate than the IFM estimates and do not benefit from increases in $d$.

\emph{Remark 4:} Occasionally, poor parameter estimates are observed for model M1 which might suggest convergence to a local maximum.  However, these estimates become much more stable with a larger sample size. This model involves 7 parameters that control both tails of the joint distribution, requiring a larger sample size to achieve accurate parameter estimates. On the other hand, this model is much more computationally stable than the skew-$t$ copula \citep{Yoshiba2018}, and it includes two parameters, $\alpha_L$ and $\alpha_U$, that independently control lower and upper tail dependence. 

\section{Real data application}
\label{sec-empstudy}

We apply the copula models based on the random processes in \eqref{eq-skewmodel1} and \eqref{eq-skewmodel2} to analyze daily mean temperatures across the state of Oklahoma, USA. The dataset spans November 1, 2023, to April 1, 2024—a total of 153 days—and includes observations from 85 monitoring stations. The data are publicly available at
\href{https://mesonet.org/}{https://mesonet.org/}.

More specifically, we consider copulas induced by the following random-process models:
\begin{itemize}
    \item[M1] $\XX(\ss) = Z^*(\ss) + \aa(\ss)\ZM_0 + \alpha_U\EE_U - \alpha_L\EE_L$, with $\aa(\ss) = \alpha_0 + \alpha_1s_1 + \alpha_2s_2$, $\alpha_0, \alpha_1, \alpha_2 \in \mathbb{R}$
    \item[M2] $\XX(\ss) = Z^*(\ss) + \alpha_0 \EE_0 + \bb(\ss) \EE_1$, with $\bb(\ss) = \exp(\beta_0 + \beta_1s_1 + \beta_2s_2)$, $\beta_0, \beta_1, \beta_2 \in \mathbb{R}$
\end{itemize}

Here, $\ss = (s_1, s_2)^{\top}$ denotes spatial coordinates; $\ZM_0$ follows a half-normal distribution; $\EE_L, \EE_U, \EE_0$, $\EE_1$ are exponential $\mathrm{Exp}(1)$ random variables; and $Z^*(\ss), \ZM_0, \EE_L, \EE_U, \EE_0, \EE_1$ are mutually independent. For both M1 and M2, we assume $Z^*(\ss)$ is a Gaussian process with the powered-exponential covariance function $$\mathrm{Cov}\left\{Z^*(\ss_1), Z^*(\ss_2)\right\} = \exp\left\{-(||\ss_1 - \ss_2||/\mu_1)^{\mu_2}\right\}\,, \quad \mu_1 > 0, \mu_2 \in (0,2].$$
Model M1 involves 7 parameters $\alpha_0, \alpha_1, \alpha_2, \alpha_L, \alpha_U, \mu_1, \mu_2$, while Model M2 involves 6 parameters $\alpha_0, \beta_0, \beta_1, \beta_2, \mu_1, \mu_2.$

The stations are geographically dispersed across various climatic zones in Oklahoma, and we anticipate that the non-stationary models M1 and M2 will offer a better fit to the data. Additionally, we examine two special cases of model M1: (1) $\alpha_0 = \alpha_1 = \alpha_2 = 0$, corresponding to the factor copula model proposed by \cite{Krupskii.Huser.ea2018} (referred to as Model M3), and (2) $\alpha_L = \alpha_U = \alpha_0 = \alpha_1 = \alpha_2 = 0$, which reduces to the standard Gaussian copula model (referred to as Model M4). Note that all four models are defined at each set of spatial coordinates $\ss$, allowing for spatial interpolation at new locations.

\subsection{Model fitting}
\label{subsec-data-fitting}

\begin{figure}[ht]
    \centering
    \includegraphics[width=.3\linewidth]{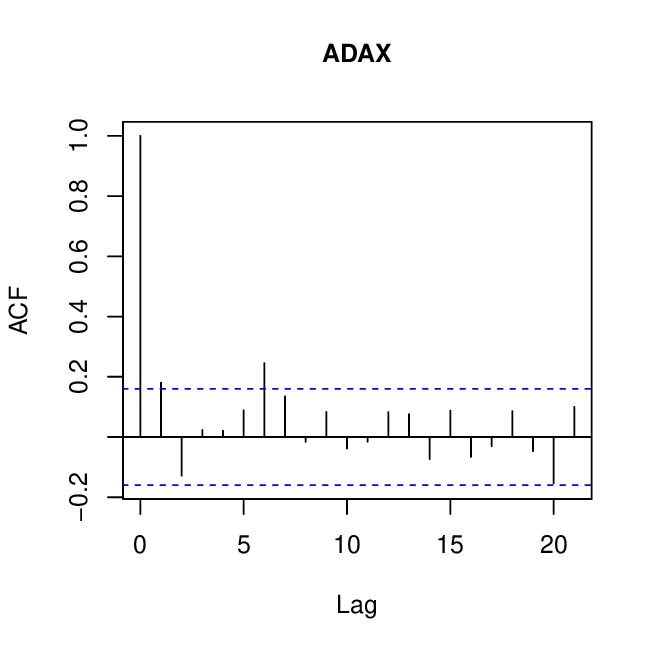}\includegraphics[width=.3\linewidth]{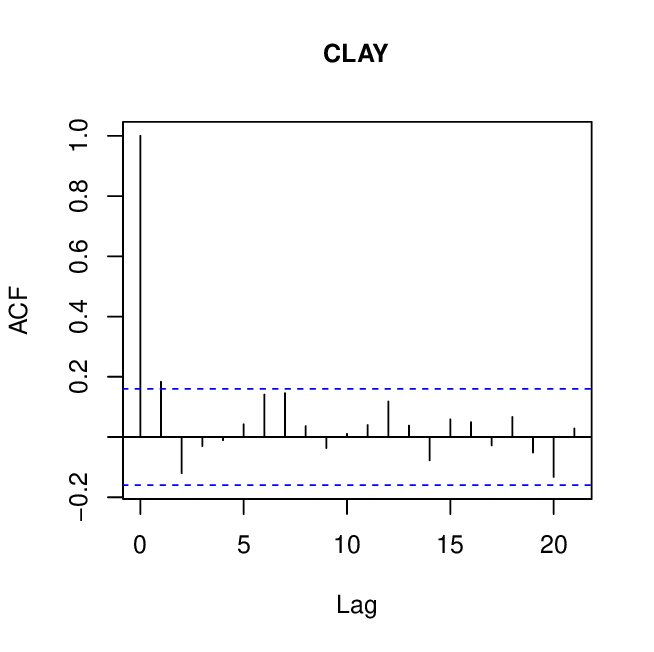}\includegraphics[width=.3\linewidth]{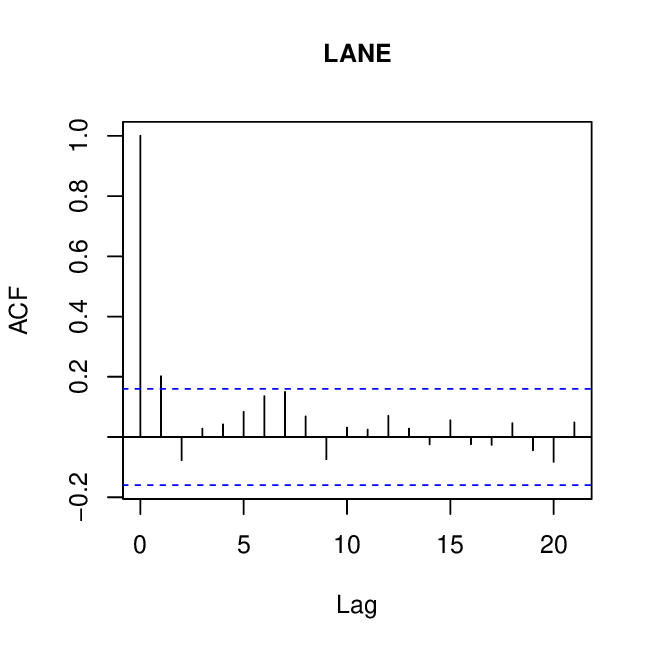}\\
    \vspace{-2mm}
    \includegraphics[width=.3\linewidth]{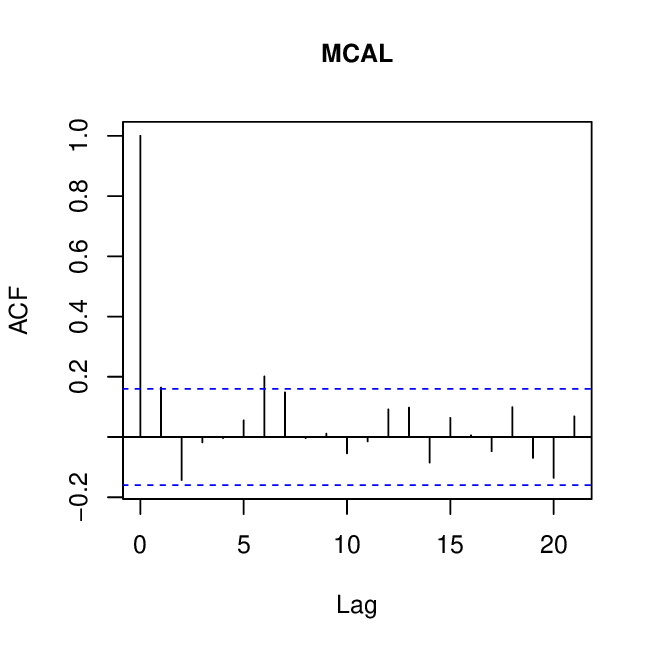}\includegraphics[width=.3\linewidth]{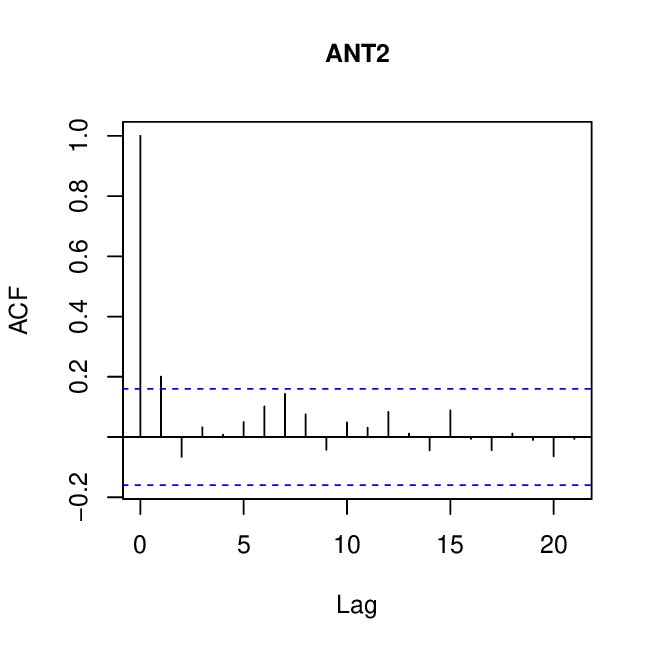}\includegraphics[width=.3\linewidth]{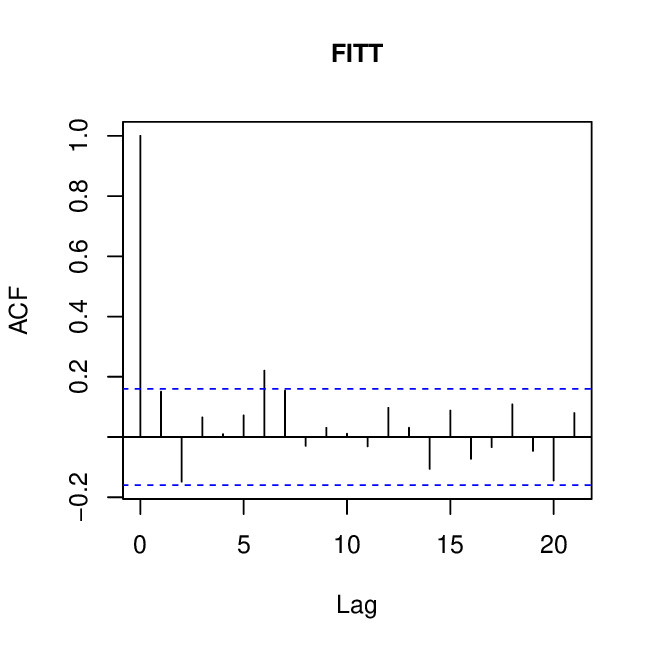}
    \vspace{-3mm}\caption{Autocorrelation plots for 6 stations with Ljung-Box test p-values below 0.01.} 
    \label{fig0}
\end{figure}

We randomly select 62 weather stations for the model fitting and 23 stations to assess the fitted model performance. For the marginals, we use the AR(2) model with spatial covariates and quadratic trend to capture temporal dependence and seasonal effect:
\begin{equation}
\label{eq-marginal}
T_{\ss,t} = c_{\ss} + c_{t} + c_1 T_{\ss, t-1} + c_2 T_{\ss, t-2} + \epsilon_{\ss,t}, \quad c_{\ss} = c_{0,\ss} + c_{1,\ss}\ss_1 + c_{2,\ss}\ss_2, \quad c_t = c_{0,t} + c_{1,t} t + c_{2,t} t^2,    
\end{equation}
where $T_{\ss,t}$ is the average daily temperature measured at location $\ss = (s_1, s_2)^{\top}$ and time $t = 1, \ldots, 153$, and  the residuals $\epsilon_{\ss,t}$ are assumed to follow the skew-$t$ distribution of \cite{Azzalini.Capitanio2003}. The marginal parameters are estimated using the maximum likelihood approach, and the data are transformed to the uniform $U(0,1)$ scale using the integral transform: $U_{\ss,t} = \widehat F_{\epsilon}(\widehat\epsilon_{\ss,t})$, where $\widehat F_{\epsilon}$ is the estimated CDF of residuals, and $\widehat\epsilon_{\ss,t}$ are the estimated residuals. Diagnostic assessments suggest that the marginal model fits the data reasonably well. Although residuals from six stations exhibit Ljung–Box test p-values below 0.01, none of the Kolmogorov–Smirnov test p-values fall below this threshold. Moreover, the autocorrelation plots for these six stations (Figure \ref{fig0}) do not show any substantial residual dependence.

\begin{figure}[ht]
    \centering
    \includegraphics[width=1\linewidth]{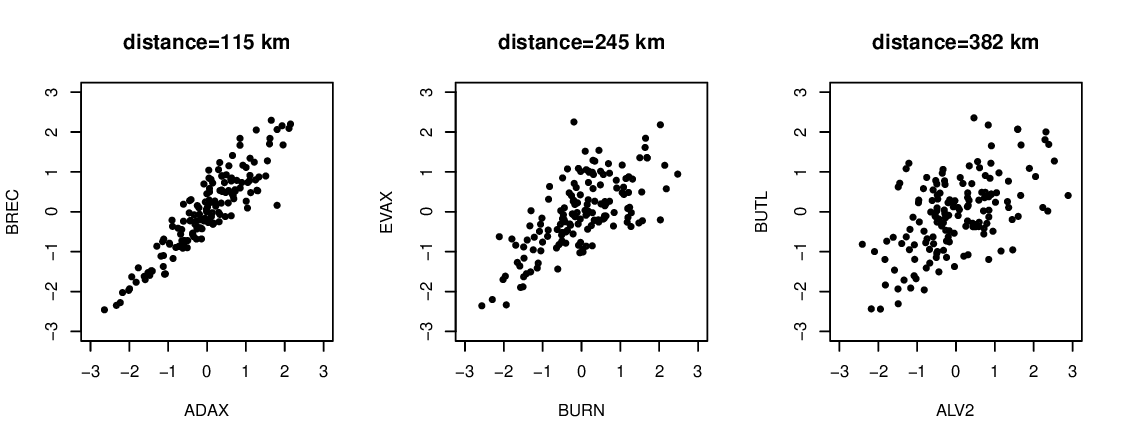}
    \caption{Normal scores scatter plots of the estimated residuals for some pairs of stations.} 
    \label{fig1}
\end{figure}

Figure \ref{fig1} shows the scatter plots  of the estimated residuals transformed to the normal $N(0,1)$ scale for some pairs of stations. Note that if the underlying copula $C_{\XX}$ linking the residuals at different locations is normal, then the elliptical shape of scatter plots is expected. However, a stronger dependence in the lower tail is observed in the data indicating that the normal copula (Model M4) might not be suitable for these data.

To further confirm these findings, we compute the lower and upper tail-weighted dependence measures introduced by \cite{Krupskii.Joe2015} for each pair $(j,k)$ of residuals $\{(U_{\ss_j, t}, U_{\ss_k, t})\}_{t=1}^{153}$, denoted $\varrho_L^{j,k}$ and $\varrho_U^{j,k}$, respectively. These measures are weighted correlations that emphasize observations near the joint tails, allowing us to quantify the strength of dependence in the lower and upper tails of a bivariate distribution. For comparison, we also calculate the same measure under the normal copula assumption (Model M1), denoted $\varrho_N^{j,k}$. Due to the symmetry of the normal copula, this value is identical in both tails. We report the average values of these measures across all pairs of weather stations: $\bar\varrho_L = 0.83$, $\bar\varrho_U = 0.59$, and $\bar\varrho_N = 0.71$. These results indicate that lower tail dependence is substantially stronger than predicted by the normal copula, whereas upper tail dependence is notably weaker.

Next, we fit Models M1, M3, M4 to the vector of residuals $\{\UU_t = (U_{\ss_1, t}, \ldots, U_{\ss_{62}, t})^{\top}\}_{t=1}^{153}$. Given the observed stronger lower tail dependence, and the fact that Model M2 is designed to capture upper tail dependence, we fit Model M2 to the reflected residuals $\{1 - \UU_t\}_{t=1}^{153}$ instead. To evaluate the fit of each model, we simulate data from the estimated models and compute model-based estimates of the Spearman correlation and the lower and upper tail-weighted dependence measures for each pair $(j,k)$ of stations, denoted by $\widehat S_{\rho}^{j,k}$, $\widehat \varrho_L^{j,k}$, and $\widehat \varrho_U^{j,k}$, respectively. These are then compared to their empirical counterparts $S_{\rho}^{j,k}$, $\varrho_L^{j,k}$, and $\varrho_U^{j,k}$, estimated from the training data.

We summarize model accuracy using the following average absolute errors over all $S = 1891$ station pairs:
$$
\Delta_{\rho} = \frac{1}{S} \sum_{j < k} |S_{\rho}^{j,k} - \widehat S_{\rho}^{j,k}|,  \quad \Delta_L = \frac{1}{S} \sum_{j < k} |\varrho_L^{j,k} - \widehat \varrho_L^{j,k}|, \quad \Delta_U = \frac{1}{S} \sum_{j < k} |\varrho_U^{j,k} - \widehat \varrho_U^{j,k}|.
$$
In addition, we compute 
$$
\Delta_3 = \frac{1}{S} \sum_{j < k} |\mu_{3}^{j,k} - \widehat \mu_{3}^{j,k}|,
$$
where $\mu_3^{j,k}$ is the empirical skewness of the difference series $\{U_{\ss_j,t} - U_{\ss_k,t}\}_{t=1}^{153}$, and $\widehat\mu_3^{j,k}$ is the corresponding model-based estimate. Under the assumption of permutation symmetry, all $\mu_3^{j,k}$ values should be close to zero.

Table~\ref{tab3} reports the values of $\Delta_\rho$, $\Delta_L$, $\Delta_U$, and $\Delta_3$, along with the BIC scores for Models M1–M4.

\begin{table}[h]
    \centering
    \begin{tabular}{l|ccccc}
     Model & $\Delta_{\rho}$ & $\Delta_L$ & $\Delta_U$ & $\Delta_3$ & BIC \\
     \hline
     M1 & 0.047 & 0.077 & 0.120 & 0.183 & $-27157$\\
     M2 & 0.059 & 0.077 & 0.095 & 0.190 & $-27125$\\
     M3 & 0.063 & 0.087 & 0.097 & 0.201 & $-26985$\\
     M4 & 0.062 & 0.133 & 0.135 & 0.201 & $-25340$\\
    \end{tabular}
    \caption{BIC values and $\Delta_{\rho}, \Delta_L, \Delta_U, \Delta_3$ for models M1-M4 fitted to the training set of 62 weather stations.}
    \label{tab3}
\end{table}

Models M1 and M2 yield lower BIC values and demonstrate better overall fit, as indicated by smaller values of $\Delta_{\rho}$. They also provide improved fit in the lower tail and more effectively capture the permutation asymmetry present in the data, as reflected by lower values of $\Delta_3$.

\subsection{Spatial interpolation}
\label{subsec-interpol}

We now use the fitted models M1–M4 to interpolate temperature values at new (testing) locations and evaluate model performance using the testing set.

Let $\ss_0$ denote a new location, and let $c_{0|1:62}$ be the conditional copula density of $U_0$ given the observed residuals at the 62 training locations, $(U_1, \ldots, U_{62})^{\top}$. Here, each $U_j$ represents the residual from the AR(2) marginal model in \eqref{eq-marginal}, transformed to have uniform $U(0,1)$ margins. For time points $t = 1$ and $t = 2$, we estimate the interpolated temperatures $\widehat T_{\ss_0,1}$ and $\widehat T_{\ss_0,2}$ by taking the average of the values from the 10 nearest neighbors. For $t \geq 3$, we compute the conditional expectation of $U_0$ given the observed residuals at time $t$ from all training locations:
\begin{equation}
\label{eq-interpol}
\E(U_{\ss_0,t}|U_{\ss_1,t},\ldots,U_{\ss_{62},t}) = \int_0^1 u c_{0|1:62}(u|U_{\ss_1,t},\ldots,U_{\ss_{62},t}) \d u = \frac{\int_0^1 u c_{0:62}(u,U_{\ss_1,t},\ldots,U_{\ss_{62},t}) \d u}{c_{1:62}(U_{\ss_1,t},\ldots,U_{\ss_{62},t})}\,,
\end{equation}
where $c_{0:62}$ and $c_{1:62}$ denote the joint copula densities of $(U_0, U_1, \ldots, U_{62})^{\top}$ and $(U_1, \ldots, U_{62})^{\top}$, respectively. Note that $c_{0,62}$ can be computed directly from the definitions of M1-M4 for each new location $\ss_0$. The integral in \eqref{eq-interpol} is computed with high accuracy using Gauss–Legendre quadrature with $n_q = 150$ points.

We next transform the conditional mean to a residual using the estimated CDF of residuals:
$$
\widehat \epsilon_{\ss_0,t} = \widehat F_{\epsilon}^{-1}\left\{\E(U_{\ss_0,t}|U_{\ss_1,t},\ldots,U_{\ss_{62},t})\right\},
$$
and use equation \eqref{eq-marginal} along with the estimated residual to interpolate the temperature at location $\ss_0$:
$$
\widehat T_{\ss_0,t} = c_{\ss_0} + c_t + c_1\widehat T_{\ss_0,t-1}
 + c_2\widehat T_{\ss_0,t-2} + \widehat \epsilon_{\ss_0,t}.$$
This procedure is repeated for $t = 3, \ldots, 153$ and for each of the 23 testing locations $\ss_0$. To assess predictive accuracy, we compute the root mean squared error (RMSE) at each location, excluding the first 10 days:
$$
\text{RMSE}_{\ss_0} = \left\{\frac{1}{143}\sum_{t=11}^{153} (T_{\ss_0, t} - \widehat T_{\ss_0, t})^2 \right\}^{1/2}\,.
$$

The overall RMSEs for the four models are 1.158 (M1), 1.162 (M2), 1.167 (M3), and 1.217 (M4), indicating that Models M1–M3 outperform M4. While the RMSEs for M1, M2, and M3 are relatively close, M1 and M2 provide significantly better predictions during extremely cold days (days 75–78), with RMSEs of 0.729, 0.702, 0.795, and 1.699, respectively. This improvement is attributed to the stronger ability of Models M1 and M2 to capture lower tail dependence. Model M4, in contrast, substantially underestimates lower tail dependence, resulting in poorer interpolation performance. Additionally, the nonstationary models M1 and M2 perform notably better at the westernmost testing station, Boise City, which is geographically isolated from the other stations. The respective RMSEs for this station are 1.916 (M1), 1.970 (M2), 2.084 (M3), and 2.249 (M4).

\begin{figure}[t!]
    \centering
    \includegraphics[width=1\linewidth]{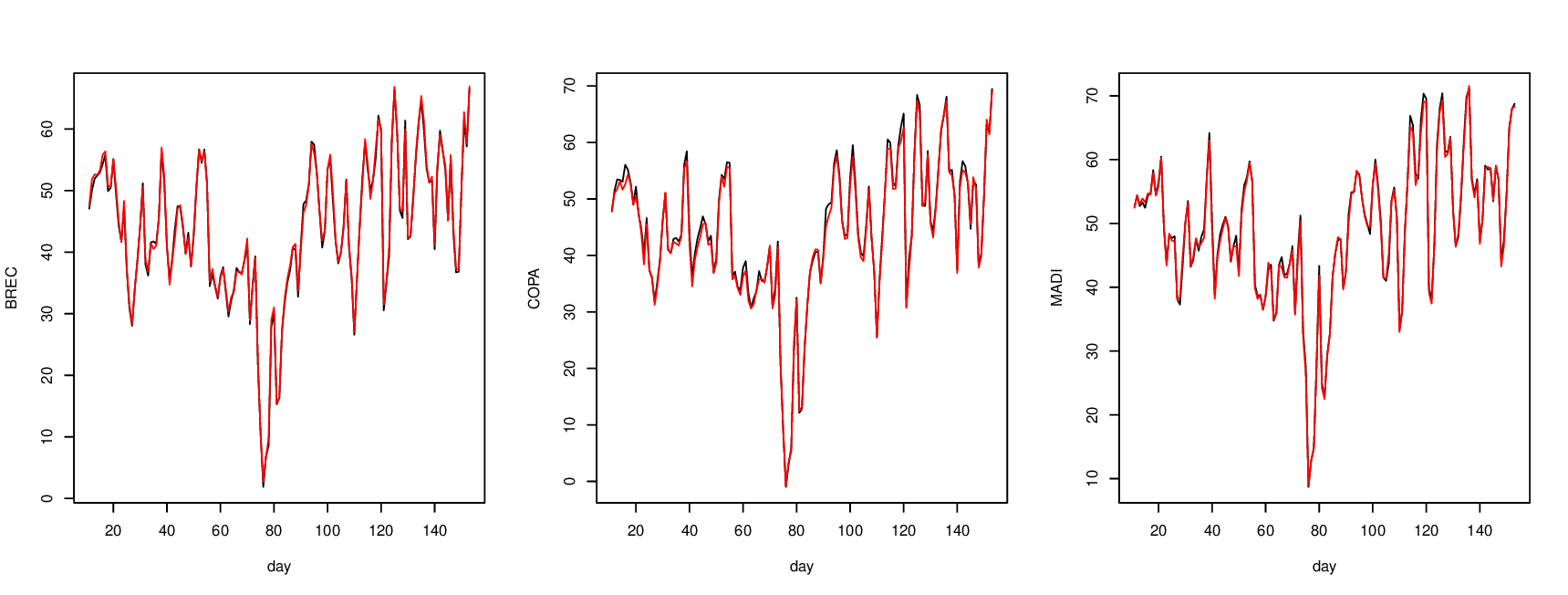}
    \caption{Daily mean temperatures recorded at three testing locations (black lines) and respective interpolated values obtained using model M1 (red lines).} 
    \label{fig2}
\end{figure}

Figure~\ref{fig2} presents the daily mean temperatures at three testing locations, along with the corresponding interpolated values obtained using Model M1. The results demonstrate excellent interpolation accuracy. Overall, Models M1 and M2 perform very well, achieving slightly lower RMSEs compared to Model M3. Moreover, M1 and M2 show substantial improvements over M3 on days with extremely low temperatures, as well as at a remote station lacking nearby neighbors.

\section{Conclusion}
\label{sec-conc}

In this paper, we explored a class of copula models derived from Pareto-mixture random processes. These models offer a parsimonious yet flexible framework for capturing both the central behavior and tail characteristics of multivariate distributions. Notably, they can accommodate tail dependence and permutation asymmetry, making them well-suited for complex spatial data. Several special cases admit closed-form likelihoods, enabling efficient maximum likelihood estimation and spatial interpolation even in high dimensions. Simulation studies demonstrate that these models are significantly more computationally stable than the skew-$t$ copula model.

Future research directions include extending the proposed class of models to multivariate data with spatial or spatio-temporal dependence, as well as conducting a more detailed investigation of their behavior at subasymptotic levels. Another promising avenue is to enhance the Pareto-mixture construction in \eqref{main-eq} by incorporating additive Pareto components. This extension would yield a richer family of distributions with greater flexibility in both the bulk of the multivariate distribution and its tails.

\appendix
\section{Appendix}
\subsection{Proof of Proposition \ref{prop3}}
\label{appx-prop3}

We have $\Pr(\ZZ > 0) = 1$ with $\bar F_{Z_i}(z) = o(z^{-\nu})$ as $z \to \infty$ for any $\nu > 0$. The density of $P$ is given by $f_P(z) = z^{1/\beta-1}/(\beta+1)$ for $z \leq 1$ and $f_P(z) = z^{-2}/(\beta+1)$ for $z > 1$, so that it satisfies assumptions of Proposition \ref{prop1} and Corollary \ref{corol1} which implies $(X_i, X_k)^{\top}$ exhibits upper tail dependence.

We will now use the following lemma to compute $\ell_{\XX}(\ww)$.

%\subsection{Tail dependence function for skew-multivariate distributions}
%\label{appdx1}

%In this section, we 

\begin{lemma}
\label{lemma1}
Consider a special case of the model \eqref{main-eq} and $\ZZ = \exp(\ZZ^* + \boldsymbol{\alpha}\vv)$ where $\vv$ is a continuous random variable with the PDF $f_{\vv}$.
Under the assumptions of Proposition \ref{prop1}, the stable tail dependence function $\ell_{\XX}$~is
    \begin{equation}
        \label{eq-skew-tailfun}
        \ell_{\XX}(w_1,\ldots,w_d) = \sum_{l = 1}^d w_l\int_{\mathbb{R}} \Phi_{ \OO_l^*}\left\{\frac{\bm\Psi_{l,-l}}{2} + \frac{1}{\bm\Psi_{l,-l}}\ln\frac{w_l^*}{\ww_{-l}^*} + \frac{v(\alpha_l\nu_l - {\bm \alpha_{-l}}{\bm \nu_{-l}})}{\bm \Psi_{l,-l}}\right\} f_{\vv}^l(v) dv,
        %\ell_{X_i,X_k}(w_i,w_k) = \sum_{l \in \{i,k\}} w_l\int_{\mathbb{R}} \Phi\left\{\frac{\Psi_{ik}}{2} + \frac{1}{\Psi_{ik}}\ln\frac{w_l^*}{w_{-l}^*} + \frac{v(\alpha_l\nu_l - \alpha_{-l}\nu_{-l})}{\Psi_{ik}}\right\} f_{\vv}^l(v) dv, 
    \end{equation}
where $\OO_l^*$ is a $(d-1) \times (d-1)$ correlation matrix with entries
$$
(\OO^*_l)_{i,k} = \frac{\Psi_{il}^2 + \Psi_{kl}^2 - \Psi_{ik}^2}{2\Psi_{il}\Psi_{kl}}\,, \quad i \neq k, \quad  i, k \in \{1,\ldots,d\} \backslash \{l\}. 
$$
Here, $w_l^* = w_l/\xi_l$ and
$$f_{\vv}^l(v) = \xi_l^{-1} \exp(\alpha_l\nu_lv)f_{\vv}(v), \quad \xi_l = \int_{\mathbb{R}}\exp(\alpha_l\nu_lv)f_{\vv}(v) \d v, \quad l \in \{1, \ldots, d\}.$$
\end{lemma}

\emph{Proof:} We use the integration by parts formula to find:
\begin{align*}
\zeta_l =& \int_0^{\infty}
 \left\{1 - F_{Z_l}(y^{1/\nu_l})\right\}\d y 
 = \int_{\mathbb{R}}\int_0^{\infty} \left\{1 - \Phi\left(\nu_l^{-1}\ln y - \alpha_l v\right)\right\}\d y f_{\vv}(v) \d v\\
 =& \int_{\mathbb{R}}
 \int_{\mathbb{R}} \left\{1 - \Phi\left(\nu_l^{-1}z - \alpha_l v\right)\right\}\exp(z)\d z f_{\vv}(v) \d v = \xi_l\exp(0.5\nu_l^2). %\int_{\mathbb{R}}\exp(\alpha_l\nu_lv) f_{\vv}(v) \d v.
 \end{align*}
 %Let $\Phi_{\Omega_{i,k}}$ be the CDF of standard bivariate normal distribution with correlation $\Omega_{i,k}$. 
 We can write:
 $$
 \ell_{\XX}(w_1,\ldots,w_d) = \int_{\mathbb{R}}g_{1:d}(v) f_{\vv}(v) \d v, $$
 and we apply the integration by parts formula to find:
 \begin{align*}
 g_{1:d}(v) :=& \int_0^{\infty} \left\{1 - \Phi_{\OO}\left({\bm\nu^{-1}}\ln y + {\bm \nu^{-1}}\ln \frac{\bm w}{\bm \zeta} - {\bm \alpha} v\right)\right\} \d y\\
 =& \int_{\mathbb{R}} \left\{1 - \Phi_{\OO}\left({\bm \nu^{-1}}z + {\bm \nu^{-1}}\ln \frac{\bm w}{\bm \zeta} - {\bm \alpha} v\right)\right\} \exp(z)\d z\\
 =& \sum_{l =1}^d \frac{w_l}{\zeta_l}\int_{\mathbb{R}} \Phi_{\tilde\OO_l}\left\{\frac{\frac{1}{\bm \nu_{-l}}\ln\frac{w_l/\zeta_l}{{\bm w_{-l}}/{\bm\zeta_{-l}}} + (\OO_{l,-l}\alpha_l - {\bm \alpha_{-l}})v + \left(\frac{\nu_{l}}{\bm \nu_{-l}} - \OO_{l,-l}\right)z}{(1-\OO_{l,-l}^2)^{1/2}} \right\}\\
 & \times \phi(z - \alpha_l v)\exp(z \nu_l) \d z \\ 
 =& \sum_{l=1}^d \frac{w_l}{\xi_l}\exp(\alpha_l\nu_lv)\int_{\mathbb{R}}\Phi_{\tilde\OO_l}\left\{\frac{ \ln\frac{w_l^*}{\bm w_{-l}^*} +  z\left(\nu_l- \OO_{l,-l}{\bm \nu_{-l}}\right) + \frac{\bm \Psi_{l,-l}^2}{2} + v(\alpha_l\nu_l - {\bm\alpha_{-l}}{\bm \nu_{-l}})}{{\bm \nu_{-l}}(1 - \OO_{l,-l}^2)^{1/2}}\right\}\phi(z) \d z\\
 =& \sum_{l =1}^d \frac{w_l}{\xi_l}\exp(\alpha_l\nu_lv)\Phi_{\OO_l^*}\left\{\frac{\bm \Psi_{l,-l}}{2} + \frac{1}{\bm\Psi_{l,-l}}\ln\frac{w_l^*}{\bm w_{-l}^*} + \frac{v(\alpha_l\nu_l - {\bm \alpha_{-l}}{\bm \nu_{-l}})}{\bm \Psi_{l,-l}}\right\}\,.
% g_{1:d}(v) :=& \int_0^{\infty} \left\{1 - \Phi_{\Omega_{i,k}}\left(\nu_i^{-1}\ln y + \nu_i^{-1}\ln \frac{w_i}{\zeta_i} - \alpha_i v, \nu_k^{-1}\ln y  + \nu_k^{-1}\ln \frac{w_k}{\zeta_k} - \alpha_k v\right)\right\} \d y\\
% =& \int_{\mathbb{R}} \left\{1 - \Phi_{\Omega_{i,k}}\left(\nu_i^{-1}z + \nu_i^{-1}\ln \frac{w_i}{\zeta_i} - \alpha_i v, \nu_k^{-1}z + \nu_k^{-1}\ln \frac{w_k}{\zeta_k}- \alpha_k v\right)\right\} \exp(z)\d z\\
% =& \sum_{l \in \{i,k\}} \frac{w_l}{\zeta_l}\int_{\mathbb{R}} \Phi\left\{\frac{\frac{1}{\nu_{-l}}\ln\frac{w_l/\zeta_l}{w_{-l}/\zeta_{-l}} + (\Omega_{i,k}\alpha_l - \alpha_{-l})v + \left(\frac{\nu_{l}}{\nu_{-l}} - \Omega_{i,k}\right)z}{(1-\Omega_{i,k}^2)^{1/2}} \right\} \phi(z - \alpha_l v)\exp(z \nu_l) \d z \\
% =& \sum_{l \in \{i,k\}} \xi_l^{-1} w_l \Phi\left\{\frac{\Psi_{ik}}{2} + \frac{1}{\Psi_{ik}}\ln\frac{w_l/\xi_l}{w_{-l}/\xi_{-l}} + \frac{v(\alpha_l\nu_l - \alpha_{-l}\nu_{-l})}{\Psi_{ik}}\right\}\exp(\alpha_l\nu_lv). 
\end{align*}
Here $\tilde\OO_l$ denotes the partial correlation matrix of $\ZZ^*$ obtained by conditioning on $Z^*_l$. The proof is now complete. \hfill$\Box$

To complete the proof of Proposition \ref{prop3}, we denote $$I_l: = \int_{\mathbb{R}} \Phi_{\OO_l^*}\left\{\frac{\bm \Psi_{l,-l}}{2} + \frac{1}{\bm \Psi_{l,-l}}\ln\frac{w_l^*}{\bm w_{-l}^*} + \frac{v(\alpha_l\nu_l - {\bm \alpha_{-l}}{\bm \nu_{-l}})}{\bm \Psi_{l,-l}}\right\} f_{\vv}^l(v) dv, \quad l \in \{1,\ldots, d\}.$$

%We now consider two examples where $I_l$ and hence $\ell_{X_i,X_k}$ are available in a simple form.
%\emph{Example 1:} 

Let $\vv$ follow the truncated normal distribution with $f_{\vv}(v) = \phi(v)/\Phi(v)$ for $v > -\tau$, then $\xi_l = \exp(0.5\alpha_l^2\nu_l^2)\Phi(\tau+\alpha_l\nu_l)/\Phi(\tau)$ and $f_{\vv}^l(v) = \phi(v - \alpha_l\nu_l)/\Phi(\tau+\alpha_l\nu_l)$ for $v > -\tau$. Following the notation from Section \ref{sec-spec-case},
\begin{align*}
I_l =& \Phi(\tau+\alpha_l\nu_l)^{-1}\int_{-\tau}^{\infty} \Phi_{\OO_l^*}\left\{\frac{\bm \Psi_{l,-l}}{2} + \frac{1}{\bm \Psi_{l,-l}}\ln\frac{w_l^*}{{\bm{\tilde w}}_{-l}^*} + \frac{v(\alpha_l\nu_l - {\bm \alpha_{-l}}{\bm \nu_{-l}})}{\bm \Psi_{l,-l}}\right\} \phi(v - \alpha_l \nu_l) dv\\
=& \Phi(\tau+\alpha_l\nu_l)^{-1}\int_{-\tau - \alpha_l\nu_l}^{\infty}
\Phi_{\OO_l^*}\left\{\frac{1}{\bm \Psi_{l,-l}} \ln \frac{\tilde w_l}{\bm{\tilde w}_{-l}}+ \frac{v(\alpha_l\nu_l - {\bm \alpha_{-l}}{\bm \nu_{-l}})}{\bm \Psi_{l,-l}} + \frac{\bm{\lambda}_{l,-l}^2}{2\bm \Psi_{l,-l}}\right\}\phi(v) \d v\\
&=\Phi(\tau+\alpha_l\nu_l)^{-1}\Phi_{\bar\OO_l^\circ}\left\{\frac{\bm{\lambda}_{l,-l}}{2} + \frac{1}{\bm{\lambda}_{l,-l}} \ln \frac{\tilde w_l}{\bm{\tilde w}_{-l}} , \tau+ \alpha_l\nu_l\right\}\,,
%\Phi(\tau+\alpha_l\nu_l)^{-1}\Phi_{\Omega_{l}^*}\left(\frac{\lambda_{ik}}{2} + \frac{1}{\lambda_{ik}} \ln \frac{\tilde w_l}{\tilde w_{-l}}, \tau+\alpha_l\nu_l\right) = \Phi_1\left(\frac{\lambda_{ik}}{2} + \frac{1}{\lambda_{ik}} \ln \frac{\tilde w_l}{\tilde w_{-l}}; 0, 1, \alpha_l^*, \tau^*_l\right),
\end{align*}
where $\bar \OO_{l}^{\circ}$ is a $d \times d$ correlation matrix with entries
$$
(\bar\OO^{\circ}_l)_{i,k} = \frac{\lambda_{il}^2 + \lambda_{kl}^2 - \lambda_{ik}^2}{2\lambda_{il}\lambda_{kl}}\,, \quad (\bar\OO^{\circ}_l)_{i,d+1} = \frac{\alpha_l\nu_l - \alpha_i\nu_i}{\lambda_{il}}\,, \quad i \neq k, \quad  i, k \in \{1,\ldots,d\} \backslash \{l\}. 
$$
It follows that $I_l$ is the CDF of a multivariate extended skew-normal distribution \cite{Beranger.Padoan.ea2019} with correlation matrix $\OO_l^{\circ}$, shape-parameter vector $\bm\alpha_l^{\circ}$ and extension parameter $\tau^{\circ}_l$. From \eqref{eq-skew-tailfun}, we then obtain  $\ell_{\XX}(w_1,\ldots,w_d) = \sum_{l=1}^dw_l I_l$ which completes the proof. \hfill $\Box$

\subsection{Proof of Proposition \ref{prop4a}}
\label{appx-prop4a}

%\emph{Example 2:}
We have $\Pr(\ZZ > 0) = 1$ and 
$$\bar F_{Z_i}(z) = \Phi(-\ln z) + \Phi(\ln z  - 1/\alpha_i)\cdot \exp(0.5/\alpha_i^2) \cdot z^{-1/\alpha_i} = O(z^{-1/\alpha_i}), \quad z \to \infty.$$  
From Proposition \ref{prop1} and Corollary \ref{corol1}, the pair $(X_i, X_k)^{\top}$ exhibits upper tail dependence whenever $1/\alpha_i > \nu_i$ and $1/\alpha_k > \nu_k$. On the other hand, one can rewrite $\XX = \ZZ P^{1/\boldsymbol{\nu}} = \exp(\ZZ^* + \boldsymbol{\nu}^{-1} \ln P) \{\exp(\EE_0)\}^{\boldsymbol{\alpha}}$, where we  define $\EE_0^* = \ln P$ (a new exponential variable), and $P^* = \exp(\EE_0)$ (a new Pareto factor). Applying Proposition \ref{prop1} and Corollary \ref{corol1} to this representation shows that $(X_i, X_k)^{\top}$ also exhibits upper tail dependence when $\alpha_i > 1/\nu_i$ and $\alpha_k > 1/\nu_k$. Combining the two conditions, we conclude that $(X_i, X_k)^{\top}$ has upper tail dependence if $\alpha_i\nu_i < 1$ and $\alpha_k \nu_k < 1$ or if $\alpha_i\nu_i > 1$ and $\alpha_k \nu_k > 1$.

Now assume $\alpha_l\nu_l < 1$, $l \in \{i,k\}$. By Lemma \ref{lemma1}, $\ell_{X_i,X_k}(w_i,w_k) = w_iI_i + w_kI_k$, where $I_l$ is defined in the previous section with $d=2$. Let $\vv$ follow the exponential distribution with $f_{\vv}(v) = \exp(-v)$ for $v > 0$. Then $\xi_l = 1/(1-\alpha_l\nu_l)$ and $f_{\vv}^l(v) = (1-\alpha_l\nu_l)\exp\left\{-(1-\alpha_l\nu_l)v\right\}$ for $v > 0$. The term $I_l$ can be calculated via the identity:
$$
\int_0^{\infty} \Phi(a + b v) \exp(-v) \d v = \Phi(a) + \begin{cases}
    \exp(a/b + 0.5/b^2)\Phi(-a-1/b), & b > 0,\\
    -\exp(a/b+0.5/b^2)\Phi(a+1/b), & b < 0,
\end{cases}
$$
with
$$a = \frac{\Psi_{ik}}{2} + \frac{1}{\Psi_{ik}}\ln \frac{\tilde w_l}{\tilde w_{-l}}\,, \quad b = (\alpha_l\nu_l - \alpha_{-l}\nu_{-l})/\left\{\Psi_{ik}(1-\alpha_l\nu_l)\right\}\,.$$

To compute $\ell_{X_i,X_k}(w_i,w_k)$ when $\alpha_l\nu_l > 1$ for $l \in \{i,k\}$, we can use the alternative representation 
$\XX = \exp(\ZZ^* + \boldsymbol{\nu}^{-1}\EE_0^*)(P^*)^{\boldsymbol{\alpha}}$. 
In this case, Lemma \ref{lemma1} remains applicable after swapping the roles of the parameters, that is, by replacing $\alpha_l$ with $1/\nu_l$ and $\nu_l$ with $1/\alpha_l$. Finally, Proposition \ref{prop2} implies that $(X_i, X_k)^{\top}$ exhibits no upper tail dependence when either $\alpha_i  < 1/\nu_i$ and $\alpha_k  > 1/\nu_k$ or  $\alpha_i  < 1/\nu_i$ and $\alpha_k > 1/\nu_k$. \hfill $\Box$

\subsection{Proof of Proposition \ref{prop4b}}
\label{appx-prop4b}

Denote by $\lambda_L^{W_i,W_k}$ the lower tail dependence coefficient of the pair $(W_i, W_k)^{\top}$. For $l \in \{i,k\}$, write $Y_l = \alpha_l\EE_0 + \beta_l\EE_1$. Let $w_l(q)$ denote the $q$-quantile of the distribution of $W_l$. Then
\begin{align*}
    \lambda_L^{W_i,W_k} & = \lim_{q \to 0} \frac{\Pr\left\{W_i = Z_i^* + Y_i < w_i(q), W_k = Z_k^* + Y_k< w_k(q)\right\}}{q} \\
    & \leq \sum_{l \in \{i,k\}} \lim_{q \to 0} \frac{\Pr\left\{Z_i^* + Y_l < w_l(q), Z_k^* + Y_l< w_l(q)\right\}}{q} = 0,
\end{align*}
since the pair $(Z_i^* + Y_l, Z_k^* + Y_l)^{\top}$ has no lower tail dependence \citep{Krupskii.Huser.ea2018}. \hfill $\Box$

\bibliographystyle{model2-names}
\bibliography{ref_ev.bib}

\end{document}